\documentclass[12pt,fleqn]{article}
\usepackage{graphicx}
\usepackage{amsfonts}

\usepackage{color}
\usepackage{graphics}
\usepackage{epsfig}
\input{psfig}

\newcommand{\R}{\mathbb R}
\renewcommand{\r}{\smallsize \mathbb R}

\newcommand{\ra}{\rightarrow}
 \newcommand{\ket}{\rangle}
\newcommand{\be}{\begin{equation}}
\newcommand{\ee}{\end{equation}}
\newcommand{\bea}{\begin{eqnarray}}
\newcommand{\eea}{\end{eqnarray}}
\newcommand{\eps}{\epsilon}

\newcommand{\ep}{\qquad {\vrule height 10pt width 8pt depth 0pt}}

\newcommand{\grintl}{[\kern-.18em [}
\newcommand{\grintr}{]\kern-.18em ]}
\newcommand{\ds}{\displaystyle}
\newcounter{smalllist}
\newenvironment{SL}{\begin{list}{{\bf\arabic{smalllist}.}}{%
\setlength{\topsep}{0mm}\setlength{\parsep}{0mm}\setlength{\itemsep}{0mm}%
\setlength{\labelwidth}{2em}\setlength{\leftmargin}{2em}\usecounter{smalllist}%
}}{\end{list}}

\newtheorem{theorem}{Theorem}[section]
\newtheorem{thm}[theorem]{Theorem}

\newtheorem{prop}[theorem]{Proposition}

\def\smallR{\hbox{\scriptsize I\kern-.23em{R}}}
\def\R{\hbox{$\mit I$\kern-.33em$\mit R$}}
\def\C{\hbox{$\mit I$\kern-.6em$\mit C$}}
\def\un{\hbox{$\mit I$\kern-.77em$\mit I$}}
\def\0{\hbox{$\mit I$\kern-.70em$\mit O$}}
\def\r{I\kern-.277em R}

\def\N{{\mathbb N}}

\begin{document}

\title{A Mathematical Theory for Vibrational Levels Associated
with Hydrogen Bonds\\ II\,:\quad The Non--Symmetric Case}

\author{George A. Hagedorn\thanks{Partially
Supported by National Science Foundation
Grant DMS--0600944.}\\
Department of Mathematics and\\
\hspace{-29pt}Center for Statistical Mechanics, Mathematical Physics,
and Theoretical Chemistry\\
Virginia Polytechnic Institute and State University\\
Blacksburg, Virginia 24061--0123, U.S.A.\\[15pt]
\and
Alain Joye\\
Institut Fourier\\ Unit\'e Mixte de Recherche CNRS--UJF 5582\\
Universit\'e de Grenoble I,\quad
BP 74\\
F--38402 Saint Martin d'H\`eres Cedex, France
}

%\author{George A. Hagedorn\thanks{Partially
%Supported by National Science Foundation
%Grant DMS--0303586.}
%\thanks{Department of Mathematics \ and \
%Center for Statistical Mechanics and Mathematical Physics,
%Virginia Polytechnic Institute and State University,
%Blacksburg, Virginia 24061-0123, U.S.A.} \and
%Alain Joye \thanks{
%Institut Fourier, UMR CNRS-UJF 5582,
%Universit\'e de Grenoble I,
%BP 74,
%38402 Saint-Martin-d'H\`eres, France }
%\thanks{Laboratoire de Physique et Mod\'elisation des Milieux Condens\'es,
%UMR CNRS-UJF 5493, Universit\'e de Grenoble I, BP 166, 38042
%Grenoble, France}}

%\date{}
\maketitle

\vskip 6mm
\begin{abstract}
We propose an alternative to the usual time--independent
Born--Oppenheimer approximation that is specifically designed
to describe molecules with non--symmetrical hydrogen bonds.
In our approach, the masses of the hydrogen nuclei are scaled
differently from those of the heavier nuclei, and we employ
a specialized form for the electron energy level surface.
As a result, the different vibrational modes appear at different
orders of approximation.

Although we develop a general theory, our analysis is
motivated by an examination of the $F\,H\,Cl^-$ ion.
We describe our results for it in detail.

We prove the existence of quasimodes and quasienergies for the nuclear
vibrational and rotational motion to arbitrary order in the Born--Oppenheimer
parameter $\eps$. When the electronic motion is also included, we provide
simple formulas for the quasienergies up to order $\eps^3$ that compare well
with experiment and numerical results.
\end{abstract}

\newpage
\baselineskip=20pt  % Alain prefers 15pt.

\section{Introduction}
\setcounter{equation}{0}

This is the second in a series of articles devoted
to the study of vibrational levels associated with
hydrogen bonds. The first paper \cite{hagjoy10} deals with
stretching vibrations of the hydrogen bond in
the symmetric case in which the hydrogen binds
two identical atoms or molecules. Our prototypical example
is $FHF^-$, which displays strong anharmonic effects,
coupling between vibrational modes, and a low frequency
for the vibration of the hydrogen along the $F$--$F$ axis.
This second paper deals with all the vibrations and
rotations in the non--symmetric situation.
Our canonical example is $F\,H\,Cl^-$, which displays weaker
anharmonic effects and a high frequency for the vibration
of the hydrogen along the $F$--$Cl$ axis.

Both of our papers contain two main new ideas. The first
is the same for both papers. Standard Born--Oppenheimer
approximations keep the electron masses fixed while all the
nuclear masses are taken proportional to $\eps^{-4}$. We
take the hydrogen mass proportional to $\eps^{-3}$ while
keeping the heavier atoms' masses proportional to $\eps^{-4}$.
This is physically appropriate for many molecules of interest:
If the mass of an electron is $1$ and $\eps$ is defined so
the mass of a carbon $C^{12}$ nucleus is $\eps^{-4}$, then
$\eps\,=\,0.0821$,and the mass of a $H^1$ nucleus is
$1.015\,\eps^{-3}$.

The second novel idea is to exploit the smallness of certain
derivatives of the electron energy level surface for the molecule
being studied. Here our two papers are completely different,
and they are motivated by examinations of numerically
computed electron energy level surfaces using Gaussian 2003 software
\cite{g03}.
In the symmetric case, the second derivative associated with moving
the $H$ along the axis of $A\,H\,A$ is small, and we could
allow it to be small and
negative if the $H$ nucleus felt a double well potential.
In the non--symmetric case, if the $H$ is more weakly bound
to the $B$ in $A\,H\,B$, we assume all the derivatives associated
with moving the $B$ relative to $A\,H$ in $A\,H\,B$ are small.
We assume all derivatives associated with stretching the distance
between $A$ and $H$ not to be small.

To describe the smallness of the small derivatives, we could have
introduced another small parameter. Instead, we have elected
to let $\eps$ play a second role. We take all the small derivatives
to be proportional to $\eps$. For the choice of $\eps=0.0821$
indicated above, that is again appropriate for our
$F\,H\,F^-$ and $F\,H\,Cl^-$ examples. The small derivatives
are on the order of $\eps$ in units where the non--small derivatives
are on the order of 1.

We shall now restrict our attention to triatomic non--symmetrical
hydrogen bonded molecules $A\,H\,B$, and assume the $H$ is more
strongly bound to the $A$. We do an asymptotic expansion for small
$\eps$, and our main results are the following:
\begin{SL}
\item To their respective leading orders, the vibrational
levels are described by three independent harmonic oscillators
in appropriate Jacobi coordinates:\quad
two separate one--dimensional harmonic oscillators
and one two--dimensional isotropic harmonic oscillator.
This is in contrast to the usual Born--Oppenheimer theory in which
one obtains one coupled four--dimensional harmonic oscillator. Our
technique does not require going through the diagonalization process
to separate the normal modes. The different modes appear at different
orders of the expansion, in contrast to the Born--Oppenheimer
situation, where all vibrations are of order $\eps^2$.
\item The highest frequency vibrational states have energy of order
$\eps^{3/2}$. These are
the stretching oscillations of the
$A$--$H$ bond with the $B$ approximately sitting still.
\item The next highest frequency vibrations are the two
degenerate bending modes. They are of order $\eps^2$.
\item The lowest vibrational energies are of order
$\eps^{5/2}$. They are the stretching oscillations of the weak
bond between the $A\,H$ and the $B$.
\end{SL}

\vskip 5mm
For the specific case of $F\,H\,Cl^-$, we have the following
comparison of results, where vibrational energies are measured in
$\mbox{cm}^{-1}$. The experimental results come from \cite{experiment}.
We note that the experiments were not done in the ``gas phase,''
so they may not accurately represent results for the isolated ions.
All the Gaussian 2003 results presented in this paper are obtained
by using the MP2 technique with the aug-cc-pvdz basis set.
The software implements the standard Born--Oppenheimer
approximation. The results for our model come from approximating
the ground state electron energy surface with Gaussian 2003
and then applying our techniques.

$$
\begin{array}{cccc}
\mbox{\bf Mode}&\mbox{\bf Experiment}\quad &\quad
\mbox{\bf Gaussian '03}\quad &\quad
\mbox{\bf Our Model}\\[3mm]
F-H\ \mbox{stretch}&2710&2960&2960\\[2mm]
\mbox{bends (degenerate)}&843&875&871\\[2mm]\qquad
FH-Cl\ \mbox{stretch}\qquad&275&246&251
\end{array}
$$

\newpage \noindent
{\bf Remarks}
\begin{SL}
\item
It is not surprising that the results for our model are close to
those obtained by Gaussian since we have used the same electron
energy surface. The Gaussian software deals with the full
4--dimensional harmonic oscillator, whereas our technique deals
with two 1--dimensional harmonic oscillators and one
isotropic 2--dimensional harmonic oscillator. Evidently the
Jacobi coordinates we have chosen are very close to the normal
mode coordinates for the 4--dimensional oscillator.\\
\item The results from Gaussian and our model are just leading
order (harmonic) calculations. Including higher order terms from the
expansions might bring these into better agreement with experiment.
Also, we again emphasize that the experimental results were not
obtained for isolated ions.
\end{SL}

\vskip 5mm
A recent chemistry article \cite{science} contains
data for vibrations of eighteen hydrogen bonded molecules
in the gas phase. It also contains an idea for quantifying
how symmetric or non--symmetric a hydrogen bond is.
Its conclusions are consistent with the analysis in our
two papers.
Figure 2 of that article plots the vibrational frequency of
the $A-H$ stretch versus the difference in
the ``proton affinities''
of $A$ and $B$ for a molecule $A\,H\,B$. When $A$ and $B$
are identical, the frequency is low (800--1000 $\mbox{cm}^{-1}$),
and when they attract the proton very differently,
the frequency is high (1600--3500 $\mbox{cm}^{-1}$).
In our symmetric analysis, this vibrational energy is of order
$\eps^2$, whereas in our non--symmetric analysis, it is of order
$\eps^{3/2}$, which is roughly 3.5 times larger when $\eps=0.0821$.

\vskip 1cm \noindent
{\bf Remarks}
\begin{SL}
\item We assume that the ground state
electron energy level we are considering is non--degenerate
for all nuclear configurations of interest.
Thus, we do not consider situations that exhibit the Renner--Teller
effect \cite{Renner,Yarkony,Herman}.
\item Since our analysis includes rotations of the whole molecule,
some small effects show up in the calculations. For example,
$l$--type doubling \cite{doubling}
occurs for terms that have non--zero eigenvalues
of the $L_{z'}$ operator at low order. ($L_{z'}$ is the
nuclear angular momentum around the $A-B$ axis.)
States corresponding to
$L_{z'}$ eigenvalue $\pm k$ with $k\ge 1$
generically have their degeneracy in energy
split at order $\eps^{2+3k}$ in our model.
\end{SL}

\newpage
The paper is organized as follows:\, In Section \ref{Sect2},
we describe our model in detail. In Section \ref{Sect3}, we do
the semiclassical expansion to all orders for the nuclei.
In Section \ref{Sect4} we include the electrons.
However, when we include the electrons, we just show that the
energy expansion is valid through order $\eps^3$. Going to
higher order is extremely complicated.
%However,
%proving that the expansion exists to all orders is so complicated
%that we simply show that the energy expansion is valid up to an
%error of order $\eps^3$.

%%%%%%%%%%%%%%%%%%%%%%%%%%%%%%%%%%%%%%%%%%%%%%
%%%%%%%%%%%%%%%%%%%%%%%%%%%%%%%%%%%%%%%%%%%%%%
\section{Semiclassical Analysis for the Effective Nuclear\\
Hamiltonian}\label{Sect2}
\setcounter{equation}{0}
%%%%%%%%%%%%%%%%%%%%%%%%%%%%%%%%%%%%%%%%%%%%%%
%%%%%%%%%%%%%%%%%%%%%%%%%%%%%%%%%%%%%%%%%%%%%%

In this section, we give a precise description of the
Hamiltonian for the nuclei.
As mentioned above, we consider a molecular system
$A\,H\,B$ in which the hydrogen is much more tightly bound
to the $A$ than to the $B$.

%
%
%
%
%%%%    Picture Stuff From Alain  %%%%%%%%%

We construct the coordinate system we use in two steps,
as illustrated in the figures below.
The first step is to choose a standard
Jacobi coordinate system for the nuclei
in their center of mass frame of reference.
The first three coordinates are the components
$X_1$, $X_2$, and $X_3$ of the vector $\vec X$ from the $A$ nucleus
to the $H$ nucleus.
The fourth, fifth, and sixth coordinates
$Y_1$, $Y_2$, and $Y_3$ are the components of the vector $\vec Y$ from
the center of mass of the $A$ and $H$ nuclei to the $B$ nucleus.
\vspace{.5cm} \\ 
\centerline{\begin{picture}(0,0)%
\includegraphics{jacobi.pstex}%
\end{picture}%
\setlength{\unitlength}{2486sp}%
\begingroup\makeatletter\ifx\SetFigFontNFSS\undefined%
\gdef\SetFigFontNFSS#1#2#3#4#5{%
  \reset@font\fontsize{#1}{#2pt}%
  \fontfamily{#3}\fontseries{#4}\fontshape{#5}%
  \selectfont}%
\fi\endgroup%
\begin{picture}(7751,6664)(2757,-7319)
\put(6931,-1186){\makebox(0,0)[lb]{\smash{{\SetFigFontNFSS{12}{14.4}{\rmdefault}{\mddefault}{\updefault}{\color[rgb]{0,0,0}{\large \bf B}}%
}}}}
\put(2926,-4246){\makebox(0,0)[lb]{\smash{{\SetFigFontNFSS{12}{14.4}{\rmdefault}{\mddefault}{\updefault}{\color[rgb]{0,0,0}{\large \bf A}}%
}}}}
\put(4636,-826){\makebox(0,0)[lb]{\smash{{\SetFigFontNFSS{7}{8.4}{\rmdefault}{\mddefault}{\updefault}{\color[rgb]{0,0,0}{\large $z$}}%
}}}}
\put(3376,-7036){\makebox(0,0)[lb]{\smash{{\SetFigFontNFSS{7}{8.4}{\rmdefault}{\mddefault}{\updefault}{\color[rgb]{0,0,0}{\large $x$}}%
}}}}
\put(8551,-4336){\makebox(0,0)[lb]{\smash{{\SetFigFontNFSS{12}{14.4}{\rmdefault}{\mddefault}{\updefault}{\color[rgb]{0,0,0}{\large \bf H}}%
}}}}
\put(9496,-5011){\makebox(0,0)[lb]{\smash{{\SetFigFontNFSS{7}{8.4}{\rmdefault}{\mddefault}{\updefault}{\color[rgb]{0,0,0}{\large $y$}}%
}}}}
\put(5086,-5101){\makebox(0,0)[lb]{\smash{{\SetFigFontNFSS{12}{14.4}{\rmdefault}{\mddefault}{\updefault}{\color[rgb]{0,0,0}{\large \bf CM of AH}}%
}}}}
\put(6481,-3661){\makebox(0,0)[lb]{\smash{{\SetFigFontNFSS{7}{8.4}{\rmdefault}{\mddefault}{\updefault}{\color[rgb]{0,0,0}{\large $\vec{X}$}}%
}}}}
\put(5761,-2401){\makebox(0,0)[lb]{\smash{{\SetFigFontNFSS{7}{8.4}{\rmdefault}{\mddefault}{\updefault}{\color[rgb]{0,0,0}{\large $\vec{Y}$}}%
}}}}
\end{picture}%
} \vspace{.5cm} 
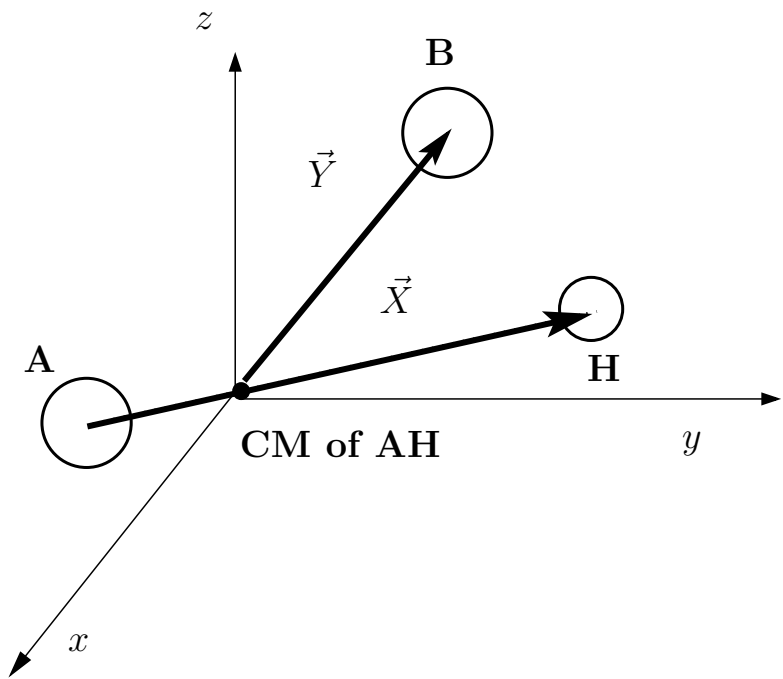
\centerline{Jacobi coordinates for the molecule}

\newpage
We now change from these coordinates to new ones that we call
$(Y,\,\theta,\,\phi,\,R,\,\gamma,\,X)$.
The $(Y,\,\theta,\,\phi)$ are spherical coordinates for the
vector described by $(Y_1,\,Y_2,\,Y_3)$ in the original
center of mass frame of reference.
The $(R,\,\gamma,\,X)$ are cylindrical coordinates for the
vector $(X_1,\,X_2,\,X_3)$ in a frame of reference that rotates
so that the axis for these coordinates is in the direction
of the vector described by $(Y_1,\,Y_2,\,Y_3)$.
The precise definition is below.
\vspace{.5cm} \\ 
\centerline{\begin{picture}(0,0)%
\includegraphics{defcoord.pstex}%
\end{picture}%
\setlength{\unitlength}{2486sp}%
\begingroup\makeatletter\ifx\SetFigFontNFSS\undefined%
\gdef\SetFigFontNFSS#1#2#3#4#5{%
  \reset@font\fontsize{#1}{#2pt}%
  \fontfamily{#3}\fontseries{#4}\fontshape{#5}%
  \selectfont}%
\fi\endgroup%
\begin{picture}(7751,7618)(2757,-7823)
\put(6931,-1186){\makebox(0,0)[lb]{\smash{{\SetFigFontNFSS{12}{14.4}{\rmdefault}{\mddefault}{\updefault}{\color[rgb]{0,0,0}{\large \bf B}}%
}}}}
\put(2926,-4246){\makebox(0,0)[lb]{\smash{{\SetFigFontNFSS{12}{14.4}{\rmdefault}{\mddefault}{\updefault}{\color[rgb]{0,0,0}{\large \bf A}}%
}}}}
\put(4636,-826){\makebox(0,0)[lb]{\smash{{\SetFigFontNFSS{7}{8.4}{\rmdefault}{\mddefault}{\updefault}{\color[rgb]{0,0,0}{\large $z$}}%
}}}}
\put(9451,-2581){\makebox(0,0)[lb]{\smash{{\SetFigFontNFSS{7}{8.4}{\rmdefault}{\mddefault}{\updefault}{\color[rgb]{0,0,0}{\large $y'$}}%
}}}}
\put(9991,-4156){\makebox(0,0)[lb]{\smash{{\SetFigFontNFSS{7}{8.4}{\rmdefault}{\mddefault}{\updefault}{\color[rgb]{0,0,0}{\large $y$}}%
}}}}
\put(6571,-7711){\makebox(0,0)[lb]{\smash{{\SetFigFontNFSS{7}{8.4}{\rmdefault}{\mddefault}{\updefault}{\color[rgb]{0,0,0}{\large $x'$}}%
}}}}
\put(3376,-7036){\makebox(0,0)[lb]{\smash{{\SetFigFontNFSS{7}{8.4}{\rmdefault}{\mddefault}{\updefault}{\color[rgb]{0,0,0}{\large $x$}}%
}}}}
\put(8596,-376){\makebox(0,0)[lb]{\smash{{\SetFigFontNFSS{7}{8.4}{\rmdefault}{\mddefault}{\updefault}{\color[rgb]{0,0,0}{\large $z'$}}%
}}}}
\put(8551,-4336){\makebox(0,0)[lb]{\smash{{\SetFigFontNFSS{12}{14.4}{\rmdefault}{\mddefault}{\updefault}{\color[rgb]{0,0,0}{\large \bf H}}%
}}}}
\put(10306,-3121){\makebox(0,0)[lb]{\smash{{\SetFigFontNFSS{7}{8.4}{\rmdefault}{\mddefault}{\updefault}{\color[rgb]{0,0,0}{\large $\vec{X}$}}%
}}}}
\put(5896,-2131){\makebox(0,0)[lb]{\smash{{\SetFigFontNFSS{7}{8.4}{\rmdefault}{\mddefault}{\updefault}{\color[rgb]{0,0,0}{\large $\vec{Y}$}}%
}}}}
\put(5356,-2896){\makebox(0,0)[lb]{\smash{{\SetFigFontNFSS{7}{8.4}{\rmdefault}{\mddefault}{\updefault}{\color[rgb]{0,0,0}{\large $\theta$}}%
}}}}
\end{picture}%
} \vspace{.5cm} 
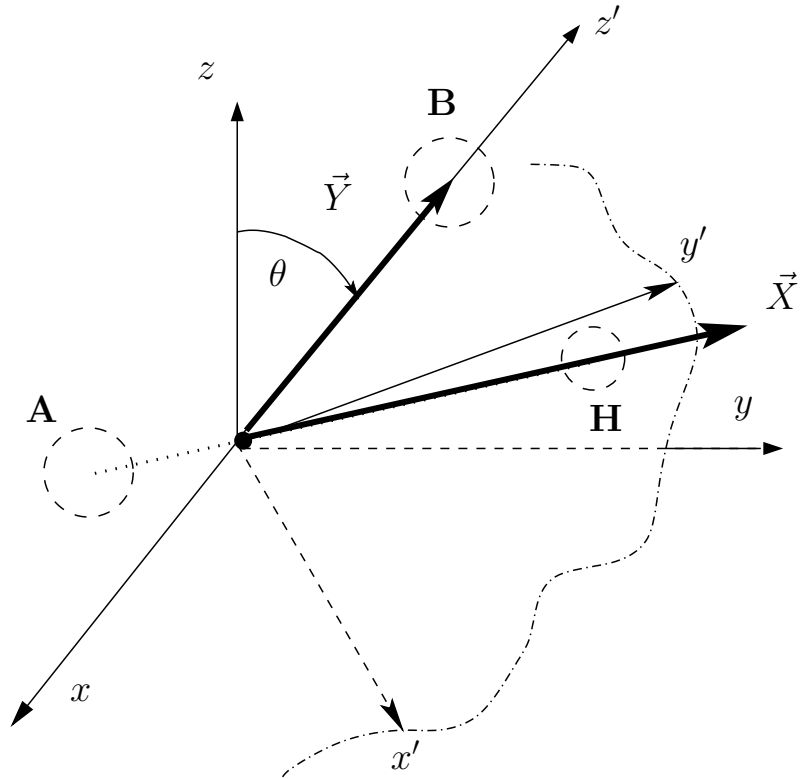
\centerline{Jacobi coordinates fixed at the origin}\\ 
\vspace{.5cm} \\ 
\centerline{\begin{picture}(0,0)%
\includegraphics{coor.pstex}%
\end{picture}%
\setlength{\unitlength}{2486sp}%
\begingroup\makeatletter\ifx\SetFigFontNFSS\undefined%
\gdef\SetFigFontNFSS#1#2#3#4#5{%
  \reset@font\fontsize{#1}{#2pt}%
  \fontfamily{#3}\fontseries{#4}\fontshape{#5}%
  \selectfont}%
\fi\endgroup%
\begin{picture}(7751,7618)(2757,-7823)
\put(7336,-6811){\makebox(0,0)[lb]{\smash{{\SetFigFontNFSS{7}{8.4}{\rmdefault}{\mddefault}{\updefault}{\color[rgb]{0,0,0}{\large $\gamma$}}%
}}}}
\put(8011,-2761){\makebox(0,0)[lb]{\smash{{\SetFigFontNFSS{7}{8.4}{\rmdefault}{\mddefault}{\updefault}{\color[rgb]{0,0,0}{\large $\gamma$}}%
}}}}
\put(4636,-826){\makebox(0,0)[lb]{\smash{{\SetFigFontNFSS{7}{8.4}{\rmdefault}{\mddefault}{\updefault}{\color[rgb]{0,0,0}{\large $z$}}%
}}}}
\put(9451,-2581){\makebox(0,0)[lb]{\smash{{\SetFigFontNFSS{7}{8.4}{\rmdefault}{\mddefault}{\updefault}{\color[rgb]{0,0,0}{\large $y'$}}%
}}}}
\put(9991,-4156){\makebox(0,0)[lb]{\smash{{\SetFigFontNFSS{7}{8.4}{\rmdefault}{\mddefault}{\updefault}{\color[rgb]{0,0,0}{\large $y$}}%
}}}}
\put(6571,-7711){\makebox(0,0)[lb]{\smash{{\SetFigFontNFSS{7}{8.4}{\rmdefault}{\mddefault}{\updefault}{\color[rgb]{0,0,0}{\large $x'$}}%
}}}}
\put(3376,-7036){\makebox(0,0)[lb]{\smash{{\SetFigFontNFSS{7}{8.4}{\rmdefault}{\mddefault}{\updefault}{\color[rgb]{0,0,0}{\large $x$}}%
}}}}
\put(8596,-376){\makebox(0,0)[lb]{\smash{{\SetFigFontNFSS{7}{8.4}{\rmdefault}{\mddefault}{\updefault}{\color[rgb]{0,0,0}{\large $z'$}}%
}}}}
\put(10306,-3121){\makebox(0,0)[lb]{\smash{{\SetFigFontNFSS{7}{8.4}{\rmdefault}{\mddefault}{\updefault}{\color[rgb]{0,0,0}{\large $\vec{X}$}}%
}}}}
\put(5896,-2131){\makebox(0,0)[lb]{\smash{{\SetFigFontNFSS{7}{8.4}{\rmdefault}{\mddefault}{\updefault}{\color[rgb]{0,0,0}{\large $\vec{Y}$}}%
}}}}
\put(5356,-2896){\makebox(0,0)[lb]{\smash{{\SetFigFontNFSS{7}{8.4}{\rmdefault}{\mddefault}{\updefault}{\color[rgb]{0,0,0}{\large $\theta$}}%
}}}}
\put(4951,-6091){\makebox(0,0)[lb]{\smash{{\SetFigFontNFSS{7}{8.4}{\rmdefault}{\mddefault}{\updefault}{\color[rgb]{0,0,0}{\large $\phi$}}%
}}}}
\put(6706,-1366){\makebox(0,0)[lb]{\smash{{\SetFigFontNFSS{7}{8.4}{\rmdefault}{\mddefault}{\updefault}{\color[rgb]{0,0,0}{\large \bf $X$}}%
}}}}
\put(7516,-5866){\makebox(0,0)[lb]{\smash{{\SetFigFontNFSS{7}{8.4}{\rmdefault}{\mddefault}{\updefault}{\color[rgb]{0,0,0}{\large \bf $R$}}%
}}}}
\end{picture}%
} \vspace{.5cm} 
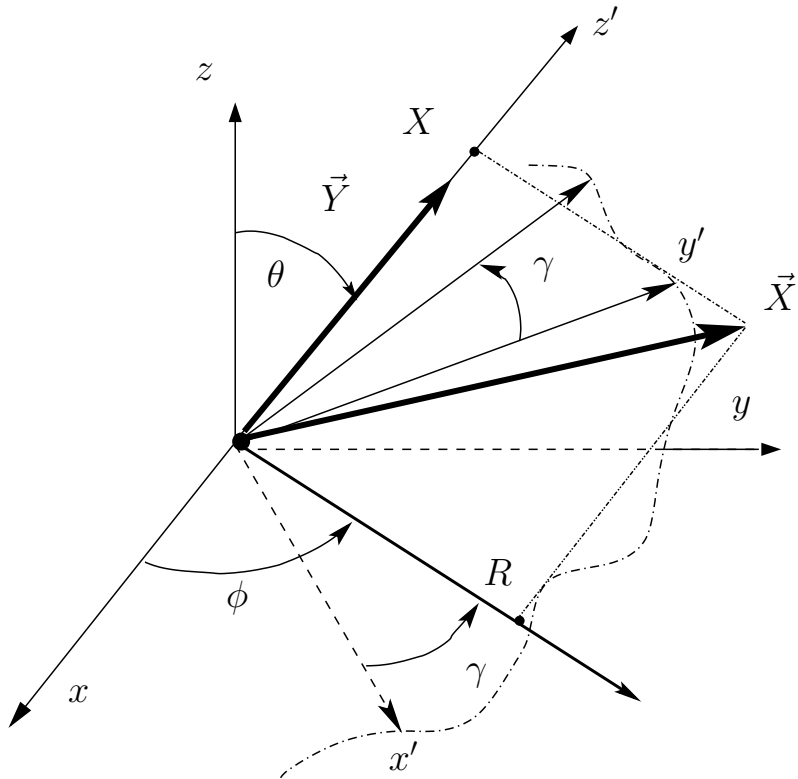
\centerline{The final coordinate system}\\

%%%%%   End Picture Stuff From Alain %%%%%%%
%
%
%
%{\large \bf ADDED FIGURE}
%\\ 
%
%
%\vskip .1cm
%We construct the coordinate system we use in two steps,
%see the figure. \vspace{.5cm} \\ 
%\centerline{\input{coor.pstex_t}} \vspace{.5cm} 
%\centerline{
%Construction of the coordinate system
%}\\
%
%The first step is to choose a standard
%Jacobi coordinate system for the nuclei
%in their center of mass frame of reference.
%The first three coordinates are the components
%$X_1$, $X_2$, and $X_3$ of the vector $\vec X$ from the $A$ nucleus
%to the $H$ nucleus.
%The fourth, fifth, and sixth coordinates
%$Y_1$, $Y_2$, and $Y_3$ are the components of the vector $\vec Y$ from
%the center of mass of the $A$ and $H$ nuclei to the $B$ nucleus.
%
%We now change from these coordinates to new ones that we call
%$(Y,\,\theta,\,\phi,\,R,\,\gamma,\,X)$.
%The $(Y,\,\theta,\,\phi)$ are spherical coordinates for the
%vector described by $(Y_1,\,Y_2,\,Y_3)$ in the original
%center of mass frame of reference.
%The $(R,\,\gamma,\,X)$ are cylindrical coordinates for the
%vector $(X_1,\,X_2,\,X_3)$ in a frame of reference that rotates
%so that the axis for these coordinates is in the direction
%of the vector described by $(Y_1,\,Y_2,\,Y_3)$.
%The precise definition is below.
%

One reason for using these coordinates is that the potential
energy surface depends only on $Y$, $X$, and $R$.
A second reason is that in these coordinates,
we can separate the total angular momentum $J^2$ and its
$z$ component $J_z$ from the other motions easily. Also,
to low order in perturbation theory, the angular momentum
$L_{z'}$ conjugate to $\gamma$ (which is the angular momentum
in the direction of $(Y_1,\,Y_2,\,Y_3)$, gives another
convenient quantum number. Note that $L_{z'}$ does not
commute with the full Hamiltonian.

The drawback to using this coordinate system is that the
kinetic energy expression is quite messy. The complication
comes from the Laplacian in the $(Y,\,\theta,\,\phi)$
variables. The Laplacian in $(R,\gamma,X)$ is simply
the usual cylindrical Laplacian.

These coordinates are closely related to ones used
in \cite{coulomb}
to deal with Born--Oppenheimer approximations for diatomic
Coulomb systems. There is a minus sign error in the expression
for $L\cdot J$ term on page 32 of that paper.

\vskip 5mm
As mentioned above, $(Y,\,\theta,\,\phi)$ are just standard
spherical coordinates. To describe the other three
coordinates precisely, we first define the rotation
$$
{\cal R}_1(\theta,\,\phi)\ =\
\left(\begin{array}{ccc}\vspace{2mm}
\cos(\theta)\,\cos(\phi)&-\,\sin(\phi)&\sin(\theta)\,\cos(\phi)
\\ \vspace{2mm}
\cos(\theta)\,\sin(\phi)&\cos(\phi)&\sin(\theta)\,\sin(\phi)
\\
-\,\sin(\theta)&0&\cos(\theta)
\end{array}\right).
$$
It maps the vector
$\left(\begin{array}{c}0\\ 0\\ 1\end{array}\right)$
to the unit vector in the direction of
$\left(\begin{array}{c}Y_1\\ Y_2\\ Y_3\end{array}\right)$.
We then define coordinates $(\xi_1,\,\xi_2,\,\xi_3)$ by
$$
\left(\begin{array}{c}\xi_1\\ \xi_2\\ \xi_3\end{array}\right)
\ =\ \left[\,{\cal R}_1(\theta,\,\phi)\,\right]^{-1}\
\left(\begin{array}{c}X_1\\ X_2\\ X_3\end{array}\right).
$$
Next, we define another rotation
$$
{\cal R}_2(\gamma)\ =\
\left(\begin{array}{ccc}\vspace{2mm}
\cos(\gamma)&-\,\sin(\gamma)&0\\ \vspace{2mm}
\sin(\gamma)&\cos(\gamma)&0\\
0&0&1
\end{array}\right),
$$
where, for generic vectors $\xi$,\ \,
$\gamma$ is defined
by requiring the second component of
$\ds\left[\,{\cal R}_2(\gamma)\,\right]^{-1}\,\xi$
to be $0$ and its first component to be positive.
We then define coordinates $X$ and $R$ by
$$
\left(\begin{array}{c}R\\ 0\\ X\end{array}\right)
\ =\ \left[\,{\cal R}_2(\gamma)\,\right]^{-1}\
\left(\begin{array}{c}\xi_1\\ \xi_2\\ \xi_3\end{array}\right).
$$

Our Hamiltonian has kinetic energy
$$
-\ \frac{\eps^3}{2\,\mu_1(\eps)}\ \Delta_{(X_1,\,X_2,\,X_3)}\
-\ \frac{\eps^4}{2\,\mu_2(\eps)}\ \Delta_{(Y_1,\,Y_2,\,Y_3)},
$$
where $\mu_1(\eps)$ and $\mu_2(\eps)$ are modified
reduced masses  that we
describe in detail below.
Since Laplacians are rotationally invariant, under our coordinate
changes, the first term simply becomes the usual cylindrical
Laplacian
$$
-\ \frac{\eps^3}{2\,\mu_1(\eps)}\ \left(\,
\frac{\partial^2\phantom{|}}{\partial R^2}\ +\
\frac 1R\ \frac{\partial\phantom{|}}{\partial R}\ +\
\frac 1{R^2}\ \frac{\partial^2\phantom{|}}{\partial \gamma^2}
\ +\ \frac{\partial^2\phantom{|}}{\partial X^2}\,\right).
$$

By a very tedious calculation,
the second term in the kinetic energy is
$$
-\ \frac{\eps^4}{2\,\mu_2(\eps)}\ \left(\,
\frac{\partial^2\phantom{|}}{\partial Y^2}\ +\
\frac 2Y\ \frac{\partial\phantom{|}}{\partial Y}\ -\
\frac 1{Y^2}\ \left\{\,J^2\,-\,2\,L\cdot J\,+\,L^2\,\right\}
\,\right),
$$
where
\be\label{totalJ}
J^2\ \,=\ \,-\
\frac{\partial^2\phantom{|}}{\partial\theta^2}\ -\
\cot\,\theta\ \frac{\partial\phantom{|}}{\partial\theta}\ -\
\frac 1{\sin^2\,\theta}\
\left(\frac{\partial^2\phantom{|}}{\partial \phi^2}+
\frac{\partial^2\phantom{|}}{\partial\gamma^2}\right)\ +\
\frac{2\,\cos\,\theta}{\sin^2\,\theta}\
\frac{\partial^2\phantom{|}}{\partial\phi\,\partial\gamma},
\ee
is the total angular momentum operator,
\bea\nonumber
L\cdot J&=&
\left(\,R\,\sin\,\gamma\
\frac{\partial\phantom{|}}{\partial X}\,-\,
X\,\sin\,\gamma\
\frac{\partial\phantom{|}}{\partial R}\,-\,
\frac XR\,\cos\,\gamma\
\frac{\partial\phantom{|}}{\partial\gamma}\right)\,
\left(\,\frac 1{\sin\,\theta}\,
\frac{\partial\phantom{|}}{\partial\phi}\,-\,
\cot\,\theta\
\frac{\partial\phantom{|}}{\partial\gamma}\right)
\\[3mm]\nonumber
&&+\quad\left(\,
R\,\cos\,\gamma\
\frac{\partial\phantom{|}}{\partial X}\,-\,
X\,\cos\,\gamma\
\frac{\partial\phantom{|}}{\partial R}\,+\,
\frac XR\,\sin\,\gamma\
\frac{\partial\phantom{|}}{\partial\gamma}\right)\,
\frac{\partial\phantom{|}}{\partial\theta}
%\ -\
%\frac{\partial^2\phantom{|}}{\partial\gamma^2},
\eea
and
$$
L^2\ =\ -\,
R^2\,\frac{\partial^2\phantom{|}}{\partial X^2}\ +\
2\,X\,R\
\frac{\partial^2\phantom{|}}{\partial X\,\partial R}\ -\
X^2\ \frac{\partial^2\phantom{|}}{\partial R^2}\ -\
\frac{X^2}{R^2}\
\frac{\partial^2\phantom{|}}{\partial\gamma^2}\ +\
\left(R-\frac{X^2}R\right)
\frac{\partial\phantom{|}}{\partial R}\ +\
2\,X\ \frac{\partial\phantom{|}}{\partial X}\ +\
\frac{\partial^2\phantom{|}}{\partial\gamma^2}.
$$

\vskip 5mm
The modified reduced masses are~
$\ds\mu_1(\eps)\,=\,\eps^3\ \frac{\eps^{-4}m_A\,\eps^{-3}m_H}
{\eps^{-4}m_A+\eps^{-3}m_H}$~
and\\
$\ds\mu_2(\eps)=\eps^4\ \frac{(\eps^{-4}m_A\,\eps^{-3}m_H)
\,\eps^{-4}m_B}
{\eps^{-4}m_A+\eps^{-3}m_H+\eps^{-4}m_B}$,~
where the three nuclei have masses $\eps^{-4}m_A$,~
$\eps^{-3}m_H$, and $\eps^{-4}m_B$.
The modified reduced masses have limits as $\eps$
tends to zero.
To isolate the leading behavior, we abuse notation
and define\ \,
$\ds \mu_1=\lim_{\eps\to 0}\,\mu_1(\eps)=m_H$\quad
and\quad
$\ds \mu_2=\lim_{\eps\to 0}\,\mu_2(\eps)=
\frac{m_A\,m_B}{m_A+m_B}$.\quad
Then we have
$$
\frac{\eps^3}{2\,\mu_1(\eps)}\ =\
\frac{\eps^3}{2\,\mu_1}\ +\
\frac{\eps^4}{2\,m_A}.
$$
Similarly,
$$
\frac{\eps^4}{2\,\mu_2(\eps)}\ =\
\frac{\eps^4}{2\,\mu_2}\ -\
\frac{\eps^5}{2\,m_A\,(m_A\,+\,2\,\eps\,m_H)}.
$$
We define the operator
$$
\eps^4\ D(\eps)\ =\
-\ \frac{\eps^4}{2\,m_A}\
\Delta_{(X_1,\,X_2,\,X_3)}\ +\
\frac{\eps^5}{2\,m_A\,(m_A\,+\,2\,\eps\,m_H)}\
\Delta_{(Y_1,\,Y_2,\,Y_3)},
$$
written in the new variables, so that
the kinetic energy can be expressed as
$$
-\ \frac{\eps^3}{2\,\mu_1}\ \Delta_{(X_1,\,X_2,\,X_3)}\
-\ \frac{\eps^4}{2\,\mu_2}\ \Delta_{(Y_1,\,Y_2,\,Y_3)}\
+\ \eps^4\ D(\eps),
$$
all written in terms of
$(Y,\,\theta,\,\phi,\,R,\,\gamma,\,X)$.

\vskip 5mm
The quantum fluctuations of the nuclei around their
equilibrium positions occur on short length scales, so
we now do the appropriate rescaling of variables. We assume
the ground state electron energy surface has a minimum
at $Y=Y_0$,~\,$R=0$~(because the Hydrogen bond is linear),
and $X=X_0$. Under the rescaling,
the angles $\theta$,~$\phi$~and $\gamma$ remain
unchanged, but we replace $Y$,~$R$,~and $X$ by
$$
y\ =\ (Y-Y_0)/\eps^{3/4},\qquad r\ =\ R/\eps^{1/2},\qquad\mbox{and}
\qquad x\ =\ (X-X_0)/\eps^{3/4}.
$$
Under this rescaling, the total kinetic energy operator becomes
\bea\label{dog}
&&\hspace{-1cm}-\ \frac{\eps^{3/2}}{2\,\mu_1}\
\frac{\partial^2\phantom{|}}{\partial x^2}\ -\
\frac{\eps^2 }{2\,\mu_1}\ \left(\,
\frac{\partial^2\phantom{|}}{\partial r^2}\ +\
\frac 1r\ \frac{\partial\phantom{|}}{\partial r}\ +\
\frac 1{r^2}\ \frac{\partial^2\phantom{|}}{\partial \gamma^2}
\,\right)\ -\
\frac{\eps^{5/2}}{2\,\mu_2}\
\frac{\partial^2\phantom{|}}{\partial y^2}
\\[3mm]\nonumber
&&-\
\frac{\eps^{13/4}}{\mu_2\,(Y_0+\eps^{3/4}y)}\
\frac{\partial\phantom{|}}{\partial y}\ +\
\frac{\eps^4}{2\,\mu_2\,(Y_0+\eps^{3/4}y)^2}\
\left\{\,J^2\,-\,2\,L\cdot J\,+\,L^2\,\right\}
\quad+\quad\eps^4\ D(\eps),
\eea
where $J^2$ is still given by (\ref{totalJ}), but
$L\cdot J$ and $L^2$ are now given by the $\eps$--dependent
expressions

\bea\nonumber\hspace{-17mm}
L\cdot J&=&
\left(\,\eps^{-1/4}\,r\,\sin\,\gamma\
\frac{\partial\phantom{|}}{\partial x}\,-\,
\eps^{-1/2}(X_0+\eps^{3/4}x)\,\sin\,\gamma\
\frac{\partial\phantom{|}}{\partial r}\,-\,\eps^{-1/2}\,
\frac{X_0+\eps^{3/4}x}{r}\,\cos\,\gamma\
\frac{\partial\phantom{|}}{\partial\gamma}\right)
\\[4mm]\nonumber
&&\hspace{90mm}
\times\quad\left(\,\frac 1{\sin\,\theta}\,
\frac{\partial\phantom{|}}{\partial\phi}\,-\,
\cot\,\theta\
\frac{\partial\phantom{|}}{\partial\gamma}\right)
\\[5mm]\nonumber
&&\hspace{-4mm}+\ \left(\,
\eps^{-1/4}r\,\cos\,\gamma\
\frac{\partial\phantom{|}}{\partial x}\,-\,
\eps^{-1/2}(X_0+\eps^{3/4}x)\,\cos\,\gamma\
\frac{\partial\phantom{|}}{\partial r}\,+\,\eps^{-1/2}
\frac{X_0+\eps^{3/4}x}{r}\,\sin\,\gamma\
\frac{\partial\phantom{|}}{\partial\gamma}\right)
\\[5mm]\nonumber
&&\hspace{13cm}\times\quad
\frac{\partial\phantom{|}}{\partial\theta}
%\\[5mm]\nonumber
%&&\hspace{-2mm}-\ \,
%\frac{\partial^2\phantom{|}}{\partial\gamma^2},
\eea
and
\bea\nonumber
L^2&=&-\
\eps^{-1/2}r^2\
\frac{\partial^2\phantom{|}}{\partial x^2}\ +\
2\ \eps^{-3/4}\ (X_0+\eps^{3/4}x)\ r\
\frac{\partial^2\phantom{|}}{\partial x\,\partial r}\ -\
\eps^{-1}(X_0+\eps^{3/4}x)^2\
\frac{\partial^2\phantom{|}}{\partial r^2}
\\[4mm]\nonumber
&&-\quad
\frac{\eps^{-1}\ (X_0+\eps^{3/4}x)^2}{r^2}\
\frac{\partial^2\phantom{|}}{\partial\gamma^2}\ +\
\eps^{-1}\,
\left(\eps\,r\,
-\,\frac{(X_0+\eps^{3/4}x)^2}{r}\right)
\ \frac{\partial\phantom{|}}{\partial r}
\\[4mm]\nonumber
&&+\quad
2\,\eps^{-3/4}\,(X_0+\eps^{3/4}x)\
\frac{\partial\phantom{|}}{\partial x}\
+\ \frac{\partial^2\phantom{|}}{\partial\gamma^2}.
\eea

\vskip 6mm
\noindent
{\bf Remarks}
\begin{SL}
\item The operator $L\cdot J$ can be rewritten
as\\[-2mm]
\bea\nonumber\hskip -5mm
L\cdot J&=&
\eps^{-1/4}\ \,\frac r2\ \,
\frac{\partial\phantom{|}}{\partial x}\ \,
\Big(\,L_{+'}\ -\ L_{-'}\,\Big)
%\\[3mm]\nonumber
%&&
\ \,-\ \,\eps^{-1/2}\ \,
\frac{X_0\,+\,\eps^{3/4}\,x}2\ \,
\frac{\partial\phantom{|}}{\partial r}\ \,
\Big(\,L_{+'}\ -\ L_{-'}\,\Big)
\\[3mm]\label{Lcoupling}
&&-\quad i\ \eps^{-1/2}\ \,
\frac{X_0\,+\,\eps^{3/4}\,x}{2\,r}\ \,
\frac{\partial\phantom{|}}{\partial\gamma}\ \,
\Big(\,L_{+'}\ +\ L_{-'}\,\Big),
\eea
where
$$L_{\pm'}\ \,=\ \,
e^{\pm i\gamma}\ \left(\,
\pm\ \frac{\partial\phantom{|}}{\partial\theta}\
+\ i\,\cot\,\theta\
\frac{\partial\phantom{|}}{\partial\gamma}\ -\ i\,
\frac 1{\sin\,\theta}\
\frac{\partial\phantom{|}}{\partial\phi}\,\right).
$$
By explicit computation, one can verify that
$L_{+'}$ and $L_{-'}$ commute with both $J^2$
and $J_z$. The operators $L_{+'}$ and $L_{-'}$
are raising and lowering operators for
the eigenstates of $L_{z'}$.
\item
The dominant order terms in the expressions in
$L\cdot J$ and $L^2$ are the ones of order
$\eps^{-1}$ in $L^2$.
Because of the overall factor of $\eps^4$ that multiplies
these operators in the Hamiltonian, they are not
relevant until the order $\eps^3$ perturbation calculations.
\end{SL}

\vskip 1cm
Motivated by numerical calculations for the $FHCl^-$ ion,
we assume the ground state electron energy
surface near its minimum depends only weakly on $R$ and $Y$.
To exploit this, we decompose the potential energy
surface as
\bea
\label{surface}V_1(X)\ +\ \eps\ V_2(X,\,R,\,Y),
\eea
where $V_1$ and $V_2$ have Taylor expansions of the forms
\bea\label{t1}
V_1(X)&\sim&
a_0\ +\ \sum_{j=2}^{\infty}\ a_j\,(X-X_0)^j,\qquad\qquad
\qquad\qquad\mbox{and}
\\[3mm]\label{t2}
V_2(X,\,R,\,Y)&\sim&
\sum_{\scriptsize
\begin{array}{c}j+k+l\ge 2\\ k+l\ge 1\\ k\ \mbox{even}\end{array}}\,
b_{j,\,k,\,l}\ (X-X_0)^j\,R^k\,(Y-Y_0)^l.
\eea
The restrictions on the indices in $V_2$ are obtained
requiring all pure $X$ dependence to be $V_1$ and by requiring
$V_2$ to be even in $R$ (because of the symmetry).

\vskip 5mm
We now can state our results for the semiclassical analysis
of the bound states for the nuclei.

\begin{thm}\label{semiclassical}
Consider the Hamiltonian
$$H(\eps)\ =\
-\ \frac{\eps^3}{2\,\mu_1(\eps)}\ \Delta_{(X_1,\,X_2,\,X_3)}\
-\ \frac{\eps^4}{2\,\mu_2(\eps)}\ \Delta_{(Y_1,\,Y_2,\,Y_3)}\
+\ V_1(X)\ +\ \eps \ V_2(X,\,R,\,Y),
$$
rewritten in terms of the variables
$(X,\,R,\,Y,\,\theta,\,\phi,\,\gamma)$.
Assume $V_1$ and $V_2$ are $C^\infty$ functions that
satisfy (\ref{t1}) and (\ref{t2}).
Assume $V_1$ has a unique global minimum $a_0$ at $X=X_0>0$,
with $a_2>0$ in (\ref{t1}), and that~
$\liminf_{|X|\to\infty}\,V(X)\,>\,a_0$.
Assume $V_2$ has a unique global minimum of~ $0$~ at~
$X=X_0$,~ $R=0$,~ and~ $Y=Y_0>0$,~ with~
$b_{0,2,0}>0$~ and~ $b_{0,0,2}>0$~ in (\ref{t2}).
Given any integer $N>0$, there exist a quasimode~
$\ds\Psi_{N/4}(\eps)\,=\,
\sum_{l=0}^N\,\eps^{l/4}\,\psi_{l/4}$~
and a quasienergy~
$\ds E_{N/4}(\eps)\,=\,
\sum_{l=0}^N\,\eps^{l/4}\,{\cal E}_{l/4}$,~
such that~ $\|\psi_{l/4}\|=O(1)$~ for each~ $l$,~
${\cal E}_{l/4}=O(1)$~ for each~ $l$,~ and 
$$
\left\|\,\Big(\,H(\eps)\,-\,E_{N/4}(\eps)\,\Big)\ 
\Psi_{N/4}(\eps)\,\right\|\quad\le\quad
C_N\ \eps^{(N+1)/4},
$$
for some~ $C_N$~ that depends on the choices of~
$n$,~ $k$,~ $m$,~ and~ $p$~ below.\\
Furthermore,
$$
{\cal E}_0\,=\,a_0,\qquad
{\cal E}_{1/4}\,=\,{\cal E}_{2/4}\,=\,{\cal E}_{3/4}
\,=\,{\cal E}_{4/4}\,=\,{\cal E}_{5/4}\,=\,{\cal E}_{7/4}
\,=\,{\cal E}_{9/4}\,=\,{\cal E}_{11/4}\,=\,0,
$$
$$
{\cal E}_{6/4}\ =\
\sqrt{\,2\,a_2/\mu_1\,}\ \left(\,
n\,+\,\frac 12\,\right),\qquad\mbox{for}\quad
n\,=\,0,\,1,\,\cdots,
$$
$$
{\cal E}_{8/4}\ =\
\sqrt{\,2\,b_{0,2,0}/\mu_1\,}\
(2\,m\,+\,|k|\,+\,1),\qquad\mbox{for an integer}\quad
k,\quad\mbox{and}\quad m\,=\,0,\,1,\,\cdots,
$$
$$
{\cal E}_{10/4}\ =\
\sqrt{\,2\,b_{0,0,2}/\mu_2\,}\ \left(\,
p\,+\,\frac 12\,\right),\qquad\mbox{for}\quad
p\,=\,0,\,1,\,\cdots,
$$
and~ ${\cal E}_{12/4}$~ is given by the expression
(\ref{E3}). The rotational energy first appears
in~ ${\cal E}_{16/4}$.\\
For fixed angular momentum quantum numbers $j$
and $j_z$, for order $N\ge 12$, the states with
$k=0$ are non-degenerate, and the states with
$|k|>0$ have multiplicity at most 2.
\end{thm}

\vskip 4mm
\noindent
{\bf Remark}\quad Theorem \ref{semiclassical} is stated with
global hypotheses and without growth conditions on the
potential. When the electronic motion is also included,
the potential energy surface may only exist locally. 
The cutoff functions that are introduced in Proposition
\ref{local} allow us to obtain analogous results with
only local assumptions.

%In that
%case, cutoff functions can be introduced to obtain the
%analogous results. See Proposition \ref{local}. 

\vskip 4mm
For the $FHCl^-$ ion, we have calculated values
for the first few coefficients in the expansion for $V$,
based on numerically differentiating results from Gaussian 2003.
Here distances are measured in Angstroms, energies are in
Hartrees, and we have used $\eps=0.0821$.
$$
\begin{array}{ccr}
a_0&=&-\,560.160\\
a_2&=&0.567\\
b_{0,2,0}&=&0.597\\
b_{1,0,1}&=&0.853\\
b_{0,0,2}&=&0.664
\end{array}
$$
%
% Data quoted are from \D:\newstuff\tex\hagjoy11\Set_UpFHCl.tex
% and \D:\newstuff\tex\hagjoy11\JacobiCoordinateNEW.nb
% Do not use numbers from \D:\newstuff\FHCLminus\2007\Fitting.nb
%

\vskip 5mm
The $\eps$ in (\ref{surface}) reflects the weakness of the
hydrogen bond, and also that the molecule can bend easily.
The $FHCl^-$ ion essentially looks like a slightly deformed
$FH$ molecule with a $Cl^-$ ion quite a long
way from the $FH$. Gaussian 2003 assigns charges associated with
each atom, and it obtains:
$$
\begin{array}{cr}
F&\quad -0.58\\
H&\quad 0.51\\
Cl&\quad -0.93\end{array}
$$
The calculated $F$--$H$ distance is 0.98 Angstrom,
and the $H$--$Cl$ distance is 1.91 Angstroms.\\
(For $HF$ alone, the charges are $\pm0.33$, the $H$--$F$ distance
is $0.925$ Angstrom, and the calculated vibrational
frequency is $4083\ \mbox{cm}^{-1}$.)

Experimental values \cite{experiment} for the vibrational
frequencies of $FHCl^-$ (in $\mbox{cm}^{-1}$)
are
%
%  I could not find where I got these numbers, so I have used some
%  earlier numbers from the literature. See the reference below.
%
%
%$$
%\begin{array}{rl}
%247&\quad FH\quad \mbox{oscillates relative to the}\quad Cl\\
%843&\quad \mbox{bends (2 degenerate modes)}\\
%2710&\quad FH\quad \mbox{oscillates}\end{array}
%$$
%
%
%  The following come from JC Evans and G Y-S Lo
%  J Phys Chem 70, 543
%  which is also referenced in
%  J Emsley "Very Strong Hydrogen Bonding"
%  www.rsc.org/ejarchive/CS/1980/CS98009000091.pdf
%
$$
\begin{array}{rl}
275&\quad FH\quad \mbox{oscillates relative to the}\quad Cl\\
843&\quad \mbox{bends (2 degenerate modes)}\\
2710&\quad FH\quad \mbox{oscillates}\end{array}
$$
Gaussian 2003 calculates the harmonic vibrational frequencies
(in $\mbox{cm}^{-1}$) to be
$$
\begin{array}{rl}
246&\quad FH\quad \mbox{oscillates relative to the}\quad Cl\\
875&\quad \mbox{bends (2 degenerate modes)}\\
2960&\quad FH\quad \mbox{oscillates}\end{array}
$$
To leading order, our model has these frequencies proportional to
$\eps^{3/2}$,\ $\eps^2$,\ and $\eps^{5/2}$ respectively.
The specific harmonic frequencies that we obtain for $FHCl^-$ are
$$
\begin{array}{rl}
251&\quad FH\quad \mbox{moves relative to the}\quad Cl\\
871&\quad \mbox{bends (2 degenerate modes)}\\
2960&\quad FH\quad \mbox{oscillates}\end{array}
$$

%%%%%%%%%%%%%%%%%%%%%%%%%%%%%%%%%%%%%%%%%%%%%%
%%%%%%%%%%%%%%%%%%%%%%%%%%%%%%%%%%%%%%%%%%%%%%
\section{The Perturbation Expansion for the Nuclei}
\label{Sect3}
\setcounter{equation}{0}
%%%%%%%%%%%%%%%%%%%%%%%%%%%%%%%%%%%%%%%%%%%%%%
%%%%%%%%%%%%%%%%%%%%%%%%%%%%%%%%%%%%%%%%%%%%%%

We now do the perturbation expansion for the
semiclassical motion of the nuclei under the global hypotheses
of Theorem \ref{semiclassical}. When the hypotheses are
satisfied only locally, see Proposition \ref{local}.
%
%Note that the $\ds J_z=-\,i\,\frac{\partial\phantom{i}}{\partial\phi}$
%commutes with the full Hamiltonian.
%As a result, we choose a value for the corresponding eigenvalue
%$j_z=0,\,\pm 1,\,\pm 2,\,\cdots$
%and henceforth assume that the $\phi$ dependence of the entire
%wave function is given by a factor of $e^{i\,j_z\,\phi}$.
%To simplify the notation, we now drop all reference to $\phi$.
%
%After removing the $\theta$ dependence, our effective
%Hamiltonian takes the form
%$$
%-\ \frac {\eps^3}2\ \frac{\partial^2\phantom{x}}{\partial X^2}
%\ -\ \frac {\eps^3}2\ \left(\,
%\frac{\partial^2\phantom{r}}{\partial R^2}\,+\,
%\frac 1R\,\frac{\partial\phantom{r}}{\partial R}\,+\,
%\frac{\lambda^2}{R^2}\,\right)\
%-\ \frac {\eps^4}2\ \frac{\partial^2\phantom{x}}{\partial Y^2}
%\ +\ V_1(X)\ +\ \eps\ V_2(X,R,Y).
%$$
%
%I really should explain, but the distinguished scaling for
%this problem is given by the following change of variables:
%$$
%x\ =\ X/\eps^{3/4},\qquad r\ =\ R/\eps^{1/2},\qquad\mbox{and}
%\qquad y\ =\ Y/\eps^{3/4}.
%$$
%Thus, w

The perturbation expansion describes the small $\eps$
dependence of  the eigenvalue problem for the following
differential operator
%\bea\nonumber&&\hspace{-1cm}
%-\ \frac {\eps^{3/2}}2\ \frac{\partial^2\phantom{x}}{\partial x^2}
%\ -\ \frac {\eps^2}2\ \left(\,
%\frac{\partial^2\phantom{r}}{\partial r^2}\,+\,
%\frac 1r\,\frac{\partial\phantom{r}}{\partial r}\,+\,
%\frac{\lambda^2}{r^2}\,\right)\
%-\ \frac {\eps^{5/2}}2\ \frac{\partial^2\phantom{x}}{\partial y^2}
%\\[4mm]\nonumber
%&&\qquad\qquad\qquad\quad +\quad a_0\
%+\ \sum_{j=2}^{\infty}\ a_j\ \eps^{3j/4}\ x^j\ +
%\sum_{\scriptsize
%\begin{array}{c}j+k+l\ge 2\\ k+l\ge 1\\ k\ \mbox{even}\end{array}}\,
%b_{j,\,k,\,l}\ \eps^{1+\frac{3(j+l)+2k}4}\ x^j\,r^k\,y^l.
%\eea
\bea\label{bigdog}
&&\hspace{-1cm}-\ \frac{\eps^{3/2}}{2\,\mu_1}\
\frac{\partial^2\phantom{|}}{\partial x^2}\ -\
\frac{\eps^2 }{2\,\mu_1}\ \left(\,
\frac{\partial^2\phantom{|}}{\partial r^2}\ +\
\frac 1r\ \frac{\partial\phantom{|}}{\partial r}\ +\
\frac 1{r^2}\ \frac{\partial^2\phantom{|}}{\partial \gamma^2}
\,\right)\ -\
\frac{\eps^{5/2}}{2\,\mu_2}\
\frac{\partial^2\phantom{|}}{\partial y^2}
\\[3mm]\nonumber
&&\hspace{24mm}-\
\frac{\eps^{13/4}}{\mu_2\,(Y_0+\eps^{3/4}y)}\
\frac{\partial\phantom{|}}{\partial y}\ +\
\frac{\eps^4}{2\,\mu_2\,(Y_0+\eps^{3/4}y)^2}\
\left\{\,J^2\,-\,2\,L\cdot J\,+\,L^2\,\right\},
\\[4mm]\nonumber
&&\qquad\qquad\qquad\quad +\quad a_0\
+\ \sum_{j=2}^{\infty}\ a_j\ \eps^{3j/4}\ x^j\ +
\sum_{\scriptsize
\begin{array}{c}j+k+l\ge 2\\ k+l\ge 1\\ k\ \mbox{even}\end{array}}\,
b_{j,\,k,\,l}\ \eps^{1+\frac{3(j+l)+2k}4}\ x^j\,r^k\,y^l.
\eea

At this point we should make the Ansatz that the eigenvalue
and eigenfunction have expansions of the forms
$$
{\cal E}\ =\ \sum_{l=0}^{\infty}\ \nu_l(\eps)\ {\cal E}_{q_l}\qquad
\mbox{and}\qquad
\psi(x,\,r,\,y,\,\theta,\,\phi,\,\gamma)\
=\ \sum_{l=0}^{\infty}\ \nu_l(\eps)\
\psi_{q_l}(x,\,r,\,y,\,\theta,\,\phi,\,\gamma).
$$
Here, $\nu_0(\eps)=1$, $\psi_0$ is non-trivial, and
$\nu_{l+1}(\eps)/\nu_l(\eps)\,\rightarrow\,0$
as $\eps\,\rightarrow\,0$.

However, one learns that every $\nu_l(\eps)$
that occurs is some power of
$\eps^{1/4}$, so it is somewhat simpler just to take
$\nu_l(\eps)=\eps^{l/4}$, {\it i.e.},
$$
{\cal E}\ =\ \sum_{l=0}^{\infty}\ \eps^{l/4}\ {\cal E}_{l/4}\qquad
\mbox{and}\qquad
\psi(x,\,r,\,y,\,\theta,\,\phi,\,\gamma)\ =\
\sum_{l=0}^{\infty}\ \eps^{j/4}\
\psi_{l/4}(x,\,r,\,y,\,\theta,\,\phi,\,\gamma).
$$

Our Hamiltonian, $J^2$, and $J_z$ all commute with one
another, so we can simultaneously diagonalize these three
operators. The eigenvalues of $J^2$ are $j(j+1)$, where
$j=0,\,1,\,2,\,\dots$, and for a given $j$, they have
degeneracy $(2j+1)^2$.
We henceforth use the specific basis for the eigenspace
for fixed $j$ that is given in Section 4.7 of \cite{Edmonds}:
$$\{\,|\,j,\,j_z,\,k\,\rangle\,:\
j_z=-j,\,-j+1,\,\dots\,,\,j;\ \,
k=-j,\,-j+1,\,\dots\,,\,j\,\},$$
where
$$
J_z\ |\,j,\,j_z,\,k\,\rangle\ =\
j_z\ |\,j,\,j_z,\,k\,\rangle\qquad\mbox{and}\qquad
L_{z'}\ |\,j,\,j_z,\,k\,\rangle\ =\
k\ |\,j,\,j_z,\,k\,\rangle,
$$
where
$\ds J_z=-\,i\,\frac{\partial\phantom{j}}{\partial\phi}$
and
$\ds L_{z'}=-\,i\,\frac{\partial\phantom{j}}{\partial\gamma}$.
Note that although $J^2$, $J_z$, and $L_{z'}$ all commute
with one another,
$L_{z'}$ does not commute with the Hamiltonian.

For future reference, we note also that the operators
in (\ref{Lcoupling}) have
$$
L_{+'}\ |\,j,\,j_z,\,k\,\rangle\ =\ \alpha_{+,j,j_z,k}\
|\,j,\,j_z,\,k+1\,\rangle\qquad\mbox{and}\qquad
L_{-'}\ |\,j,\,j_z,\,k\,\rangle\ =\ \alpha_{-,j,j_z,k}\
|\,j,\,j_z,\,k-1\,\rangle,
$$
for some $\alpha_{\pm,j,j_z,k}$.\ \, When\ \,$|k|=j$,\ \,
$\alpha_{+,j,j_z,j}=0$\ \, and\ \, $\alpha_{-,j,j_z,-j}=0$.

By restricting attention to given values of $j$ and $j_z$,
the wave functions in our expansion can now be regarded
(with some abuse of notation) as
$$
\psi_{l/4}(x,\,r,\,y,\,\theta,\,\phi,\,\gamma)\ =\
\sum_{k=-j}^j\ \psi_{l/4}(x,\,r,\,y,\,k)\
|\,j,\,j_z,\,k\,\rangle.
$$

We now substitute the Ansatz into the eigenvalue equation and
equate terms order by order.
We do not worry about normalization, but produce a quasimode that is
$O(1)$ as $\eps$ tends to $0$. To simplify some of the discussion,
we take $\psi_{l/4}$ orthogonal to $\psi_0$ for $l>0$. The
results of these computations yield the formal expansions
of Theorem \ref{semiclassical}.

\vskip 8mm \noindent
{\bf Order ${\bf \eps^{0}}$}\ \qquad
These terms simply require\ \,
$a_0\ \psi_{0}\ =\ {\cal E}_0\ \psi_{0}$.\quad So,
$$
{\cal E}_0\ =\ a_0.
$$

\vskip 6mm \noindent
{\bf Order ${\bf \eps^{l/4}}$ for ${\bf 1\le l\le 5}$}\ \qquad
The terms of these orders successively require\newline
$a_0\ \psi_{l/4}\ =\ {\cal E}_0\ \psi_{l/4}\,+\,
{\cal E}_{l/4}\ \psi_0$.\quad So,
$$
{\cal E}_{l/4}\ =\ 0.
$$

\vskip 6mm \noindent
{\bf Order ${\bf \eps^{6/4}}$}\ \qquad
These terms require\quad
$\ds -\ \frac 1{2\mu_1}\ \frac{\partial^2\psi_0}{\partial x^2}\
+\ a_2\ x^2\ \psi_0\ =\ {\cal E}_{6/4}\ \psi_0.$

\vskip 6mm \noindent
This forces
$$
{\cal E}_{6/4}\ =\ \left(n+\frac 12\right)\ \sqrt{2\,a_2/\mu_1}
\qquad\mbox{for some}\quad n\,=\,0,\,1,\,\cdots,
$$
and
$$
\psi_0(x,\,r,\,y,\,k)\ =\
f_0(r,\,y,\,k)\ \Phi_1(x),
$$
where
$$\Phi_1(x)\ =\ (2\,a_2\,\mu_1)^{1/8}\ \pi^{-1/4}\
2^{-n/2}\ (n!)^{-1/2}\ H_n(x')\ e^{-{x'}^2/2}
$$
with~ $x'=(2\,a_2\,\mu_1)^{1/4}\,x$.~
The function $f_0$ is not yet determined.

\vskip 6mm \noindent
{\bf Order ${\bf \eps^{7/4}}$}\ \qquad
We introduce the notation
$$
H_{0,x}\ =\
-\ \frac 1{2\mu_1}\ \frac{\partial^2\phantom{x}}{\partial x^2}\
+\ a_2\ x^2.
$$
Then the $\eps^{7/4}$ terms require\quad
$[H_{0,x}-{\cal E}_{6/4}]\ \psi_{1/4}\ =\ {\cal E}_{7/4}\ \psi_0$.\\
We first examine the components of this equation that are multiples
of $\Phi_1(x)$. These $\|_x$ components require
$$
{\cal E}_{7/4}\ =\ 0.
$$
We then examine the components that are perpendicular to $\Phi_1(x)$
in the $x$ variables. These $\perp_x$ components require
$$
\psi_{1/4}(x,\,r,\,y,\,k)\ =\
f_{1/4}(r,\,y,\,k)\ \Phi_1(x),
$$
where the function $f_{1/4}$ is not yet determined.

\vskip 6mm \noindent
{\bf Order ${\bf \eps^{8/4}}$}\ \qquad
These terms require
$$
[H_{0,x}-{\cal E}_{6/4}]\ \psi_{2/4}\
-\ \frac 1{2\mu_1}\
\left(\,\frac{\partial^2\psi_0}{\partial r^2}\ +\
\frac 1r\ \frac{\partial\psi_0}{\partial r}\ +\
\frac{1}{r^2}\
\frac{\partial^2\psi_0}{\partial \gamma^2}\,\right)
\ +\ b_{0,2,0}\ r^2\ \psi_0\ =\ {\cal E}_{8/4}\ \psi_0.
$$

\vskip 5mm \noindent
The $\|_x$ components of this equation require
$$
H_{0,r,\gamma}\ \psi_0\ =\ {\cal E}_{8/4}\ \psi_0,
$$
where
$$ H_{0,r,\gamma}\ =\
-\ \frac 1{2\mu_1}\
\left(\,\frac{\partial^2\phantom{r}}{\partial r^2}\ +\
\frac 1r\ \frac{\partial\phantom{r}}{\partial r}\ +\
\frac{1}{r^2}\
\frac{\partial^2\phantom{r}}{\partial \gamma^2}\,\right)
\ +\ b_{0,2,0}\ r^2.
$$
This is a standard isotropic two dimensional Harmonic
oscillator problem that one can solve
by separating variables.
In our context, the angular operator\ \,
$\ds L_{z'}\,=\,-\,i\,
\frac{\partial\psi_0}{\partial \gamma}$\ \,
has eigenvalues\ \,
$k=0,\,\pm 1,\,\pm 2,\,\cdots,\,\pm j$\ \, and
eigenfunctions\ \, $e^{i\,k\,\gamma}$.\ \,
For each such $k$, the radial operator
$$
-\ \frac 1{2\mu_1}\ \left(\,
\frac{\partial^2\phantom{r}}{\partial r^2}\ +\
\frac 1{r}\ \frac{\partial\phantom{r}}{\partial r}\ -\
\frac{k^2}{r^2}\,\right)\ +\ b_{0,2,0}\ r^2
$$
has eigenvalues
$$
{\cal E}_{8/4}\ =\
\left(\,2\,m\,+\,|k|\,+\,1\,\right)\
\sqrt{2\,b_{0,2,0}/\mu_1},\qquad
\mbox{where}\quad
m\,=\,0,\,1,\,\cdots.
$$
The corresponding normalized eigenfunctions are
$$
\sqrt{\frac{2\ (m!)}{(m+|k|)!}}\ \,
\left(\,2\,b_{0,2,0}\,\mu_1\,\right)^{1/4}\
(r')^{|k|}\ L_m^{|k|}({r'}^2)\
e^{-r'^2/2},
$$
where,\ \,$r'=(2\,b_{0,2,0}\,\mu_1)^{1/4}\,r$,\ \,
$m\ge 0$,\ \,and\ \, $L_m^{|k|}$\ \, is a Laguerre
polynomial.

We permanently fix one such value of~
${\cal E}_{8/4}$.~ 
Since different pairs $(m,\,k)$ can occur,
we define
$$
K\ =\ \{\,k\in\mathbb{Z}\,:\,|k|\le j,\
\mbox{and}~ m(k)\ge 0
\,\}.
$$
where
$$
m(k)\ =\ \frac 12\ \left(\,
{\cal E}_{8/4}\left/\sqrt{2\,b_{0,2,0}/\mu_1}\right.
\ \,-\ |k|\ -\
1\,\right).
$$

One can easily show that $K$ is non-empty and
has at most $j+1$ elements.

For~ $k\in K$,~ we define the normalized
wave functions
$$
\Phi_2(|k|,\,r)\ =\
\sqrt{\frac{2\ (m(k)!)}{(m(k)+|k|)!}}\ \,
\left(\,2\,b_{0,2,0}\,\mu_1\,\right)^{1/4}\
(r')^{|k|}\ L_{m(k)}^{|k|}({r'}^2)\
e^{-r'^2/2}
$$
and take
$$
f_0(r,\,y,\,k)\ =\
\left\{\ \begin{array}{cl}
g_0(y,\,k)\ \Phi_2(|k|,\,r)&\quad \mbox{if}\quad
k\in K\\[3mm]
0&\quad\mbox{otherwise.}\end{array}\right.
$$

The functions $g_0(y,\,k)$ for $k\in K$
are not yet determined. However, we now have
$$
\psi_0(x,\,r,\,y,\,\theta,\,\phi,\,\gamma)\ =\
\sum_{k\in K}\
g_0(y,\,k)\ \Phi_1(x)\ \Phi_2(|k|,\,r)\
|\,j,\,j_z,\,k\,\rangle.
$$

For future reference, we let $Z_1$ denote the subspace
spanned by
$$
\left\{\,
\Phi_1(x)\ \Phi_2(|k|,\,r)\
|\,j,\,j_z,\,k\,\rangle\ :\ k\in K\,\right\}.
$$

\vskip 4mm
The $\perp_x$ terms at this order require
$[H_{0,x}-{\cal E}_{6/4}]\ \psi_{2/4}\ =\ 0$,
which simply forces
$$
\psi_{2/4}\ =\ f_{2/4}(r,\,y,\,k)\ \Phi_1(x).
$$

\vskip 6mm \noindent
{\bf Order ${\bf \eps^{9/4}}$}\qquad These terms require
\bea\nonumber
&&[H_{0,x}-{\cal E}_{6/4}]\ \psi_{3/4}\ +\
[H_{0,r,\gamma}-{\cal E}_{8/4}]\ \psi_{1/4}
\ +\ a_3\ x^3\ \psi_0
\\[3mm]\label{9}
&=& {\cal E}_{9/4}\ \psi_0.
\eea
The $\|_x$ components of this equation require
\be\label{9'}
[H_{0,r,\gamma}-{\cal E}_{8/4}]\ \psi_{1/4}
\ =\ {\cal E}_{9/4}\ \psi_0.
\ee

We first examine the components of this equation that
belong to the subspace $Z_1$.
These $\|_x\,\|_{Z_1}$ components require
$$
{\cal E}_{9/4}\ =\ 0.
$$
Next, the $\|_x\perp_{Z_1}$
components of (\ref{9'}) that are orthogonal to
$Z_1$ require
$[H_{0,r,\gamma}-{\cal E}_{8/4}]\ \psi_{1/4}\,=\,0$.
This forces us to choose
$$
f_{1/4}(r,\,y,\,k)\ =\
\left\{\ \begin{array}{cl}
g_{1/4}(y,\,k)\ \Phi_2(|k|,\,r)&\quad \mbox{if}\quad
k\in K\\[3mm]
0&\quad\mbox{otherwise.}\end{array}\right.
$$

The $\perp_x$ components of (\ref{9}) require
$[H_{0,x}-{\cal E}_{6/4}]\ \psi_{3/4}\ +\ a_3\ x^3\ \psi_0\ =\ 0$.
We solve this equation by applying the reduced resolvent operator
$[H_{0,x}-{\cal E}_{6/4}]^{-1}_r$. The result is
\bea\nonumber
\psi_{3/4}(x,\,r,\,y,\,k)&=&
-\ a_3\ \sum_{k\in K}\ g_0(y,\,k)\
\Phi_2(|k|,\,r)\
[H_{0,x}-{\cal E}_{6/4}]^{-1}_r\
\left(\,x^3\ \Phi_1(x)\,\right)
\\[3mm]\label{psi34}
&&+\quad f_{3/4}(r,\,y,\,k)\ \Phi_1(x).
\eea

\vskip 6mm \noindent
{\bf Order ${\bf \eps^{10/4}}$}
\bea\nonumber
&&\hspace{-17mm}[H_{0,x}-{\cal E}_{6/4}]\ \psi_{4/4}\,+\,
[H_{0,r,\gamma}-{\cal E}_{8/4}]\ \psi_{2/4}
\\[3mm]\label{10}
&-&\frac 1{2\mu_2}\ \frac{\partial^2\psi_0}{\partial y^2}\ +\
a_3\ x^3 \psi_{1/4}\,+\,
b_{0,0,2}\ y^2\ \psi_0\,+\,b_{1,0,1}\ x\,y\,\psi_0
\quad=\quad
{\cal E}_{10/4}\ \psi_0.
\eea

The $\|_x\,\|_{Z_1}$ components require\quad
$\ds -\ \frac 1{2\mu_2}\
\frac{\partial^2\psi_0}{\partial y^2}\ +\
b_{0,0,2}\ y^2\ \psi_0\ =\
{\cal E}_{10/4}\ \psi_0$.
This forces us to choose
$$
{\cal E}_{10/4}\ =\ \left(p+\frac 12\right)\
\sqrt{2\,b_{0,0,2}/\mu_2}\qquad\mbox{where}\quad
p\,=\,0,\,1,\,\cdots,
$$
and
\be\label{g0}
g_0(y,\,k)\ =\ c_{0,k}\ \Phi_3(y)
\qquad\mbox{if}\quad k\in K,
\ee
where
$$
\Phi_3(y)\ =\ (2\,b_{0,0,2}\,\mu_2)^{1/8}\
\pi^{-1/4}\ 2^{-p/2}\ (p!)^{-1/2}\ 
H_p(y')\ e^{-{y'}^2/2}
$$
with~
$y'=(2\,b_{0,0,2}\,\mu_2)^{1/4}\,y$.

So far, the $c_{0,k}$ in (\ref{g0}) are arbitrary
for $k\in K$,
but we henceforth assume they satisfy the normalization
condition
$$
\sum_{k\in K}\ \left|\,c_{0,k}\,\right|^2\ =\ 1.
$$
For future reference, we let $Z_2$ denote the subspace
spanned by
$$
\left\{\,
\Phi_1(x)\ \Phi_2(|k|,\,r)\
\Phi_3(y)\
|\,j,\,j_z,\,k\,\rangle\ :\ k\in K\,\right\}.
$$

The $\|_x\perp_{Z_1}$ components require
$$
f_{2/4}(r,\,y,\,k)\ =\
\left\{\ \begin{array}{cl}
g_{2/4}(y,\,k)\ \Phi_2(|k|,\,r)&\quad \mbox{if}\quad
k\in K\\[3mm]
0&\quad\mbox{otherwise.}\end{array}\right.
$$

The $\perp_x$ components require\quad
$[H_{0,x}-{\cal E}_{6/4}]\ \psi_{4/4}\ +\
a_3\ x^3 \psi_{1/4}\ +\ b_{1,0,1}\ x\,y\ \psi_0\ =\ 0$.\\
We apply the reduced resolvent of $H_{0,x}$ to obtain
\bea\nonumber
\psi_{4/4}(x,\,r,\,y,\,k)&=&-\ a_3\
g_{1/4}(y,\,k)\ \Phi_2(|k|,\,r)\
[H_{0,x}-{\cal E}_{6/4}]^{-1}_r\ \left( x^3\,\Phi_1(x)\right)
\\[3mm]\nonumber
&&-\ b_{1,0,1}\ c_{0,k}\ y\ \Phi_3(y)\
\Phi_2(|k|,\,r)\
[H_{0,x}-{\cal E}_{6/4}]^{-1}_r\
\left( x^{\phantom{|}}\,\Phi_1(x)\right)
\\[3mm]\nonumber
&&+\ f_{4/4}(r,\,y,\,k)\ \Phi_1(x).
\eea
Note that the first two terms are zero if $k\notin K$.

\newpage \noindent
{\bf Remarks}
\begin{SL}
\item At this point, we have completely determined $\psi_0$,
except for the values of $c_{0,k}$ for $k\in K$.
Restoring the angular dependence in the notation, we have
$$
\psi_0\ =\ \sum_{k\in K}\
c_{0,k}\ \Phi_1(x)\ \Phi_2(|k|,\,r)\ \Phi_3(y)\ \,
|\,j,\,j_z,\,k\,\rangle.
$$
Since $j$ and $j_z$ are fixed, this is a linear combination
of at most~ $j+1$~ linearly independent states.
\item As we shall see, the degeneracy generically
partially splits at order $\eps^{12/4}$. At that point,
states with different values of $|k|$ have different
energy, but two states with $k=\pm\lambda$ for
$\lambda>0$ have the same ${\cal E}_{12/4}$.
In terms of the energy, the degeneracy of these two states
generically splits completely at
order $\eps^{2+3\lambda}$.~ When $\lambda=1$, this splitting
has long been observed in the spectra of linear polyatomic
molecules. It is called $l$--type doubling \cite{doubling}.
\item We have determined the dominant terms
for the eigenvalue:

\vskip -10pt
\bea\nonumber
&&\hskip -15mm {\cal E}_0\ +\
\eps^{3/2}\,\left(n+\frac 12\right)\,\sqrt{2\,a_2/\mu_1}\ +\
\eps^2\,\left(2m(k)+|k|+1\right)\,\sqrt{2\,b_{0,2,0}/\mu_1}
\\[3mm]\nonumber
&&\hskip 8cm +\quad
\eps^{5/2}\,\left(p+\frac 12\right)\,
\sqrt{2\,b_{0,0,2}/\mu_2}.
\eea

\vskip 10pt
This quantity does not depend on the quantum numbers
$j$, $j_z$, or $k\in K$.\\
The dominant contribution to the energy from the total angular
momentum is\\
$\ds\frac{j(j+1)\,\eps^{4^{\phantom{|}}}}{2\,\mu_2\,Y_0^2}$,~
so it enters at order $16/4$.
%However, contributions to the
%wave functions from
%$L^2_{z'}$ appear at order $12/4$ and from
%$L\cdot J$ at order $14/4$.

\vskip 10pt
\item
Below we impose the condition that every $\psi_{l/4}$
with $l>0$ be orthogonal to the subspace $Z_2$.

\item
At the next order, the pattern emerges for how to do
all higher order formal perturbation calculations. For
$l\ge 11$, we have the following:
\begin{SL}
\item[$\bullet$]
the $\|_x\,\|_{Z_1}\,\|_y$ terms determine
${\cal E}_{l/4}$,
\item[$\bullet$]
the $\|_x\,\|_{Z_1}\perp_y$ terms determine
the $y$--dependence of $g_{(l-10)/4}(y,\,k)$
%(and hence, $\psi_{(l-10)/4}(y)$ completely),
\item[$\bullet$]
the $\|_x\perp_{Z_1}$ terms determine the
$r$ and $k$ dependence of $f_{(l-8)/4}(r,\,y,\,k)$,
\quad and
\item[$\bullet$]
the $\perp_x$ terms determine the $x$--dependence of
$\psi_{(l-6)/4}(x,\,r,\,y,\,k)$.
\end{SL}
Since the general pattern occurs at the next order,
we present full calculations for only one
more order explicitly.
\end{SL}

\vskip 6mm \noindent
{\bf Order ${\bf \eps^{11/4}}$}
\bea\nonumber
&&\hspace{-25pt}[H_{0,x}-{\cal E}_{6/4}]\ \psi_{5/4}\ +\
[H_{0,r,\gamma}-{\cal E}_{8/4}]\ \psi_{3/4}\ +\
[H_{0,y}-{\cal E}_{10/4}]\ \psi_{1/4}
\\[3mm]\nonumber
&&\qquad +\
a_3\ x^3 \psi_{2/4}\ +\
b_{1,0,1}\ x\,y\ \psi_{1/4}\ +\
b_{0,2,1}\ r^2\,y\ \psi_0\ +\
b_{1,2,0}\ x\,r^2\ \psi_0
\\[3mm]\nonumber
&&=\quad
{\cal E}_{11/4}\ \psi_0.
\eea

The $\|_x\,\|_{Z_1}\,\|_y$ terms require
$$
{\cal E}_{11/4}\ =\ 0.
$$

The $\|_x\,\|_{Z_1}\perp_y$ terms require
$$
g_{1/4}(y,\,k)
\quad=\quad-\ b_{0,2,1}\ c_{0,k}\,
\left\langle\,\Phi_2(|k|,\,r),
\,r^2\,\Phi_2(|k|,\,r)\,\right\rangle_r\
[H_{0,y}-{\cal E}_{10/4}]^{-1}_r\,
\left(\,y^{\phantom{|}}\Phi_3(y)\,\right).
$$
for $k\in K$.
This is the first place in the perturbation
calculations where different values of $|k|$
yield different results.
%\\[3mm]\nonumber
%&&+\quad h_{1/4,+}(\theta)\
%e^{i\,\lambda\,\gamma}\ \Phi_3(y)
%\eea
%and
%\bea\nonumber
%g_{1/4,-}(y,\,\theta)
%&=&-\ b_{0,2,1}\ h_{0,-}(\theta)\,
%\left\langle\,\Phi_2(r),\,r^2\,\Phi_2(r)\,\right\rangle_r\
%[H_{0,y}-{\cal E}_{10/4}]^{-1}_r\
%\left(y^{\phantom{|}}\Phi_3(y)\right)
%\\[3mm]\nonumber
%&&+\quad h_{1/4,-}(\theta)\
%e^{i\,\lambda\,\gamma}\ \Phi_3(y).
%\eea
Note that we could add
$c_{1/4,k}\,\Phi_3(y)$ to $g_{1/4}(y,\,k)$ when
$k\in K$, but we have
chosen $c_{1/4,k}=0$ to
impose the condition that $\psi_{1/4}$ be
orthogonal to the subspace $Z_2$.
See Remark 4 above.

%The $\|_x\perp_{Z_1}$ terms require
%$$
%f_{3/4}(r,\,y,\,k)\ =\
%\left\{\ \begin{array}{cl}
%g_{3/4}(y,\,k)\ \Phi_2(|k|,\,r)&\quad \mbox{if}\quad
%k\in K\\[3mm]
%0&\quad\mbox{otherwise.}\end{array}\right.
%$$

The $\|_x\perp_{Z_1}$ terms require
$$
[H_{0,r,\gamma}-{\cal E}_{8/4}]\ f_{3/4}\ +\
P_{\perp_{Z_1}}\ [H_{0,y}-{\cal E}_{10/4}]\ f_{1/4}
\ +\ b_{0,2,1}\ y\ P_{\perp_{Z_1}}\ r^2\ f_0\ =\ 0,
$$
where $P_{\perp_{Z_1}}$ denotes the projection
onto functions orthogonal to the subspace
$Z_1$.
We have already seen that the non-zero
$f_{1/4}(r,\,y,\,k)$ belong to the
subspace $Z_1$, so
$P_{\perp_{Z_1}}\ [H_{0,y}-{\cal E}_{10/4}]\
f_{1/4}\ =\ 0$.
Thus, applying the reduced resolvent of
$H_{0,r,\gamma}$ (which is zero on $Z_1$),
we obtain
%\bea\nonumber
%&&f_{3/4}(r,\,y,\,k)
%\\[3mm]\nonumber
%&&\hskip -17mm
%=\quad\left\{\begin{array}{cl}
%-\quad b_{0,2,1}\
%y\ \Phi_3(y)\ %\quad\times
%\\[3mm]\nonumber
%&&
%\sum_{k'\in K}\ c_{0,k'}\
%\left\langle\,j,\,j_z,\,k\left|,\ \,
%[H_{0,r,\gamma}(|k|)-{\cal E}_{8/4}]^{-1}_r\
%P_{\perp_{Z_1}}\
%r^2\ \Phi_2(|k|,\,r)
%\ \right|\,j,\,j_z,\,k'\,\right\rangle
%\\[3mm]\nonumber
%&&\hskip -8mm
%+\ g_{3/4}(y,\,k)\ \Phi_2(|k|,\,r)&
%\mbox{if}\quad k\in K\\[3mm]
%0&\mbox{if}\quad k\notin K.
%\end{array}\right.
%\eea
\bea\nonumber
&&\hskip -16mm
f_{3/4}(r,\,y,\,k)\ =\
-\ \,b_{0,2,1}\ c_{0,k}\ y\ \Phi_3(y)\
[H_{0,r}(|k|)-{\cal E}_{8/4}]^{-1}_r\
P_{\perp_{Z_1}}\
r^2\ \Phi_2(|k|,\,r)
\\[3mm]\nonumber
&&\hskip 62mm +\quad
g_{3/4}(y,\,k)\ \Phi_2(|k|,\,r)\qquad
\mbox{if}\quad k\in K,\qquad\mbox{and}
\\[3mm]\nonumber
&&\hskip -16mm
f_{3/4}(r,\,y,\,k)\ =\quad
0\qquad\mbox{if}\quad
k\notin K.
\eea
Here, we have used the notation
$$
H_{0,r}(|k|)\ =\
-\ \frac 12\
\frac{\partial^2\phantom{r}}{\partial r^2}\
-\ \frac 1{2\,r}\
\frac{\partial\phantom{r}}{\partial r}\
+\ \frac{k^2}{2\,r^2}
$$
and the direct sum decomposition
$$
[H_{0,r,\gamma}-{\cal E}_{8/4}]^{-1}_r
\quad =\quad
\bigoplus_{|k|\le j}\quad
[H_{0,r}(|k|)-{\cal E}_{8/4}]^{-1}_r
$$
which results from $H_{0,r,\gamma}$ commuting
with $L_{z'}$.

The $\perp_x$ terms require
\bea\nonumber
&&[H_{0,x}-{\cal E}_{6/4}]\ \psi_{5/4}\ +\
[H_{0,r,\gamma}-{\cal E}_{8/4}]\ \psi_{3/4}^{\perp_x}
\\[3mm]\nonumber
&&\qquad+\quad
a_3\ x^3\ \psi_{2/4}\ +\
b_{1,0,1}\ x\,y\ \psi_{1/4}\ +\
b_{1,2,0}\ x\,r^2\ \psi_0\ =\ 0,
\eea
where $\psi_{3/4}^{\perp_x}$ denotes the component of
$\psi_{3/4}$ orthogonal to $\Phi_1(x)$ in the $x$ variables.
By combining this with
(\ref{psi34}) and (\ref{g0}), we have
$$
\psi_{3/4}^{\perp_x}(x,\,r,\,y,\,k)\
=\ \left\{\begin{array}{cl}-\ a_3\ c_{0,k}\
\Phi_3(y)\ \Phi_2(|k|,\,r)\
[H_{0,x}-{\cal E}_{6/4}]^{-1}_r\
\left(\,x^3\,\Phi_1(x)\,\right)&\quad\mbox{if}\quad
k\in K\\[3mm]
0&\quad\mbox{if}\quad k\notin K.\end{array}\right.
$$
So, we see that
$[H_{0,r}-{\cal E}_{8/4}]\ \psi_{3/4}^{\perp_x}\ =\ 0$.
Thus, we have
\bea\nonumber
\psi_{5/4}(x,\,r,\,y,\,k)&=&-\quad
a_3\ g_{2/4}(y,\,k)\
\Phi_2(|k|,\,r)\
\left(\,[H_{0,x}-{\cal E}_{6/4}]^{-1}_r\
\left(\,x^3\,\Phi_1(x)\,\right)\,\right)
\\[3mm]\nonumber
&&-\quad
b_{1,0,1}\ y\
g_{1/4}(y,\,k)\
\Phi_2(|k|,\,r)\
\left([\,H_{0,x}-{\cal E}_{6/4}]^{-1}_r\
\left(\,x\ \Phi_1(x)\,\right)\,\right)
\\[3mm]\nonumber
&&-\quad
b_{1,2,0}\ c_{0,k}\
\Phi_3(y)\ r^2\,\Phi_2(|k|,\,r)\
\left(\,[H_{0,x}-{\cal E}_{6/4}]^{-1}_r\
\left(x\ \Phi_1(x)\right)\,\right)
\\[3mm]\nonumber
&&+\quad f_{5/4}(r,\,y,\,k)\
\Phi_1(x)\hskip 58mm \mbox{if}\quad k\in K.
\eea
For $k\notin K$,
$$
\psi_{5/4}(x,\,r,\,y,\,k)\quad=\quad
f_{5/4}(r,\,y,\,k)\ \Phi_1(x).
$$
Note that only $g_{2/4}(y,\,k)$ (for $k\in K$) and
$f_{5/4}(r,\,y,\,k)$ in these expressions have not
yet been determined.

\vskip 5mm \noindent
{\bf Remarks}
\begin{SL}
\item Amazingly, $\psi_{1/4}\ne 0$.
This component of the wave function involves an anharmonic
correction related to the bending and $AH$--$B$ stretching
modes. Restoring the angular
dependence to the notation, we have
\bea\nonumber
&&\hskip -17mm \psi_{1/4}(x,\,r,\,y,\,\theta,\,\phi,\,\gamma)
\\[3mm]\nonumber
&&\hskip -17mm =\ \,-\ \,b_{0,2,1}\ \,\sum_{k\in K}\ c_{0,k}\
\left\langle\,\Phi_2(|k|,\,r),
\,r^2\,\Phi_2(|k|,\,r)\,\right\rangle_r
\\[3mm]\nonumber
&&\hskip 35mm \times\quad
\Phi_1(x)\ \Phi_2(|k|,\,r)\
[H_{0,y}-{\cal E}_{10/4}]^{-1}_r\,
\left(y^{\phantom{|}}\Phi_3(y)\right)\
|\,j,\,j_z,\,k\,\rangle.
\eea

%\bea\nonumber
%\hspace{-18.45mm}\psi_{1/4}
%&=&-\,b_{0,2,1}\,c_{0,\lambda}\,
%\left\langle\,\Phi_2(r),\,r^2\,\Phi_2(r)\,\right\rangle_r\,
%\Phi_1(x)\,\Phi_2(r)\,
%[H_{0,y}-{\cal E}_{10/4}]^{-1}_r\,
%\left(y^{\phantom{|}}\Phi_3(y)\right)\,
%|\,j,\,j_z,\,\lambda\,\rangle
%\\[3mm]\nonumber
%\hspace{-18.45mm}&&-\,b_{0,2,1}\,c_{0,-\lambda}\,
%\left\langle\,\Phi_2(r),\,r^2\,\Phi_2(r)\,\right\rangle_r\,
%\Phi_1(x)\,\Phi_2(r)\,
%[H_{0,y}-{\cal E}_{10/4}]^{-1}_r\,
%\left(y^{\phantom{|}}\Phi_3(y)\right)\,
%|\,j,\,j_z,\,-\lambda\,\rangle.
%\eea
%

\item Although we do not present the full calculations
at order $\eps^{12/4}$, we do calculate ${\cal E}_{12/4}$
explicitly. It is generically contains non-zero anharmonic
corrections.
\end{SL}

\vskip 1cm
Before going further with the expansion, we present a summary
of what has been determined so far.

\vskip 5mm
\bea\nonumber
{\cal E}&=&
{\cal E}_0\ +\
\eps^{3/2}\ \left(n+\frac 12\right)\ \sqrt{2\,a_2/\mu_1}
\quad+\quad
\eps^2\ \left(2m(k)+|k|+1\right)\ \sqrt{2\,b_{0,2,0}/\mu_1}
\\[3mm]\nonumber
&&\hspace{57mm}+\quad
\eps^{5/2}\ \left(p+\frac 12\right)\ \sqrt{2\,b_{0,0,2}/\mu_2}
\quad +\quad O(\eps^{12/4}).
\eea

\noindent
The last information for ${\cal E}$ came from order\quad
$11/4,\quad\|_x\,\|_{Z_1}\,\|_y$.

\vfill
$$
\psi_0\quad=\quad\sum_{k\in K}\
c_{0,k}\ \Phi_1(x)\ \Phi_2(|k|,\,r)\ \Phi_3(y)\ \,
|\,j,j_z,k\,\rangle.
$$
This was completely determined at order\quad
$10/4,\ \,\|_x\,\|_{Z_1}$.

\vfill
\bea\nonumber
\hspace{-9mm}\psi_{1/4}
&=&-\ b_{0,2,1}\ \sum_{k\in K}\
c_{0,k}\
\left\langle\,\Phi_2(|k|,\,r),
\ r^2\,\Phi_2(|k|,\,r)\,\right\rangle_r
\\[3mm]\nonumber
&&\hskip 51mm\times\quad
\Phi_1(x)\ \Phi_2(r)\
[H_{0,y}-{\cal E}_{10/4}]^{-1}_r\,
\left(y^{\phantom{|}}\Phi_3(y)\right)\,
|\,j,j_z,k\,\rangle
\eea
This was completely determined at order\quad
$11/4,\ \,\|_x\,\|_{Z_1}\perp_y$.

\vfill
$$
\psi_{2/4}\quad=\quad\sum_{k\in K}\
g_{2/4}(y,\,k)\
\Phi_1(x)\
\Phi_2(|k|,\,r)\
|\,j,j_z,k\,\rangle.
$$
The last information came from order
$10/4,\quad\|_x\perp_{Z_1}$.

\vfill
\bea\nonumber
\psi_{3/4}&=&
-\quad b_{0,2,1}\ \sum_{k\in K}\
c_{0,k}\
\Phi_1(x)\ \left(y\ \Phi_3(y)\right)
\\[2mm]\nonumber
&&\hskip 4cm\times\quad
[H_{0,r}(|k|)-{\cal E}_{8/4}]^{-1}_r\
\left(P_{\perp_{Z_1}}\ r^2\
\Phi_2(|k|,\,r)\right)\
|\,j,j_z,k\,\rangle
\\[4mm]\nonumber
&&+\quad\sum_{k\in K}\
g_{3/4}(y,\,k)\
\Phi_1(x)\ \Phi_2(|k|,\,r)\
|\,j,j_z,k\,\rangle.
\eea

\noindent
The last information came from order
$11/4,\quad\|_x\perp_{Z_1}$.
\pagebreak

\vfill
\bea\nonumber
\psi_{4/4}&=&a_3\ b_{0,2,1}\
[H_{0,y}-{\cal E}_{10/4}]^{-1}_r\
\left(y^{\phantom{|}}\Phi_3(y)\right)\
[H_{0,x}-{\cal E}_{6/4}]^{-1}_r\
\left( x^3\,\Phi_1(x)\right)
\\[3mm]\nonumber
&&\hspace{29mm}\times\quad
\sum_{k\in K}\
c_{0,k}\
\left\langle\,\Phi_2(|k|,\,r),\
r^2\,\Phi_2(|k|,\,r)\,\right\rangle_r\
\Phi_2(|k|,\,r)
|\,j,j_z,k\,\rangle
\\[3mm]\nonumber
&&-\quad b_{1,0,1}\ \left(\,y\ \Phi_3(y)\,\right)\
[H_{0,x}-{\cal E}_{6/4}]^{-1}_r\
\left( x^{\phantom{1}}\Phi_1(x)\right)\
\sum_{k\in K}\
c_{0,k}\
\Phi_2(|k|,\,r)\
|\,j,j_z,k\,\rangle
\\[2mm]\nonumber
&&+\quad \sum_{k=-j}^j\ f_{4/4}(r,\,y,\,k)\ \Phi_1(x)\
|\,j,j_z,k\,\rangle.
\eea

\noindent
The last information came from order
$10/4,\quad\perp_x$\\
(coupled with $11/4,\quad\|_x\,\|_{Z_1}\perp_y$,\ \,
because of $g_{1/4}$).

\bea\nonumber
\hspace{-1cm}
\psi_{5/4}&=&-\quad
a_3\ \sum_{k\in K}\ g_{2/4}(y,\,k)\
\Phi_2(|k|,\,r)\
\left(\,[H_{0,x}-{\cal E}_{6/4}]^{-1}_r\
\left(x^3\,\Phi_1(x)\right)\,\right)\
|\,j,j_z,k\,\rangle
\\[3mm]\nonumber
&&-\quad
b_{1,0,1}\ b_{0,2,1}\ \sum_{k\in K}\
c_{0,k}\
\left\langle\,\Phi_2(|k|,\,r),\
r^2\,\Phi_2(|k|,\,r)\,\right\rangle_r\
\left(\,[H_{0,x}-{\cal E}_{6/4}]^{-1}_r
\left(x\,\Phi_1(x)\right)\right)
\\[1mm]\nonumber
&&\hskip 49mm\times\quad
\Phi_2(|k|,\,r)\
\left(y\ [H_{0,y}-{\cal E}_{10/4}]^{-1}_r
\left(y^{\phantom{|}}\Phi_3(y)\right)\right)
|\,j,j_z,k\,\rangle
\\[5mm]\nonumber
&&-\quad
b_{1,2,0}\ \sum_{k\in K}\
c_{0,k}\
\Phi_3(y)\ r^2\,\Phi_2(|k|,\,r)\
\left(\,[H_{0,x}-{\cal E}_{6/4}]^{-1}_r\
\left(x\ \Phi_1(x)\right)\,\right)\
|\,j,j_z,k\,\rangle
\\[3mm]\nonumber
&&+\quad \sum_{k=-j}^j\ f_{5/4}(r,\,y,\,k)\ \Phi_1(x)\
|\,j,j_z,k\,\rangle.
\eea
The last information came from order
$11/4,\quad\perp_x$.

\vskip 1cm
We now return to describing higher orders of
the perturbation expansion. We determine
${\cal E}_{12/4}$, and explicitly write the
equations that must be solved through order
$\eps^{16/4}$. That is the order at which
the angular momentum quantum number $j$ appears,
and the degeneracy due to rotations is split.

\pagebreak

\vskip 6mm \noindent
{\bf Order ${\bf \eps^{12/4}}$}
\bea\nonumber
&&\hspace{-25pt}[H_{0,x}-{\cal E}_{6/4}]\ \psi_{6/4}
\ +\ [H_{0,r,\gamma}-{\cal E}_{8/4}]\ \psi_{4/4}\ +\
[H_{0,y}-{\cal E}_{10/4}]\ \psi_{2/4}\ +\
a_3\ x^3\ \psi_{3/4}
\\[3mm]\nonumber
&&\qquad +\quad
a_4\ x^4\ \psi_0\ +\
b_{1,0,1}\ x\,y\ \psi_{2/4}\ +\
b_{0,2,1}\ r^2\,y\ \psi_{1/4}\ +\
b_{1,2,0}\ x\,r^2\ \psi_{1/4}
\\[3mm]\nonumber
&&\qquad +\quad
b_{0,4,0}\ r^4\ \psi_0\ +\
\frac{X_0^2}{2\,\mu_2\,Y_0^2}\ H_{0,r,\gamma}\ \psi_0
\\[3mm]\nonumber
&&=\quad
{\cal E}_{12/4}\ \psi_0.
\eea

  From the $\|_x\,\|_{Z_1}\,\|_\gamma$
terms, we can easily solve for ${\cal E}_{12/4}$.

\bea\nonumber
{\cal E}_{12/4}&=&-\quad a_3^2\ \,
\langle\,\Phi_1(x),\ x^3\ [H_{0,x}-{\cal E}_{6/4}]_r^{-1}\
x^3\ \Phi_1(x)\,\rangle_x
\\[3mm]\nonumber
&&+\quad
a_4\ \,
\langle\,\Phi_1(x),\ x^4\ \Phi_1(x)\,\rangle_x
\\[3mm]\nonumber
&&-\quad b_{0,2,1}^2\ \,
\langle\,\Phi_2(|k|,\,r),\
r^2\ \Phi_2(|k|,\,r)\,\rangle_r^2\quad
\langle\,\Phi_3(y),\ y\ [H_{0,y}-{\cal E}_{10/4}]_r^{-1}\
y\ \Phi_3(y)\,\rangle_y
\\[3mm]\nonumber
&&+\quad b_{0,4,0}\ \,
\langle\,\Phi_2(|k|,\,r),\
r^4\ \Phi_2(|k|,\,r)\,\rangle_r
%\langle\,\Phi_3(y),\ y^4\ \Phi_3(y)\,\rangle_y
\\[5mm]\nonumber
&&+\quad
\frac{X_0}{2\,\mu_2\,Y_0^2}\quad
\sqrt{2\,b_{0,2,0}/\mu_1}\quad(2m(k)+|k|+1)
\eea

As long as
%$b_{0,2,1}\ne 0$ or
$b_{0,4,0}\ne 0$, this expression
yields different values for different $|k|$.
To see this, first note that the factor
$$
\langle\,\Phi_2(|k|,\,r),\
r^2\ \Phi_2(|k|,\,r)\,\rangle_r^2\ =\
\left(\,
\frac{2m(k)\,+\,|k|\,+\,1}
{\sqrt{2\,b_{0,2,0}\,\mu_1}}
\,\right)^2
$$
does not depend on~ $k$,~
and the term
$$
\frac{X_0}{2\,\mu_2\,Y_0^2}\quad
\sqrt{2\,b_{0,2,0}/\mu_1}\quad
(2m(k)\,+\,|k|\,+\,1)
$$
does not depend on~ $k$.~
In fact, the only term that has non-trivial
dependence on $k$ in ${\cal E}_{12/4}$ is
$$
\langle\,\Phi_2(|k|,\,r),\
r^4\ \Phi_2(|k|,\,r)\,\rangle_r\ =\
\frac{(2+3|k|+k^2)\,+\,6\,(|k|+1)\,m(k)\,+\,6\,m(k)^2}
{2\,b_{0,2,0}\,\mu_1}
$$
We now show that different values of
$k$ yield different values of this quantity.

Let $k_1\ge 0$ and $k_2\ge 0$ be two different values
of $|k|$ that yield the same result.
Simultaneously solving
$$
(2+3k_1+k_1^2)+6(k_1+1)m(k_1)+6m(k_1)^2=
(2+3k_2+k_2^2)+6(k_2+1)m(k_2)+6m(k_2)^2
$$
and
$$2m(k_1)+k_1+1\ =\ 2m(k_2)+k_2+1$$
forces
\bea\nonumber
m(k_1)&=&(-3\,-\,5\,k_1\,+\,k_2)/6
\\[3mm]\nonumber
m(k_2)&=&(-3\,+\,k_1\,-\,5k_2)/6.
\eea
However, $m(k_1)$ and $m(k_2)$ must both be
non-negative.
There are no simultaneous non-negative solutions to
\bea\nonumber
k_2&>&3\,+\,5\,k_1
\\[3mm]\nonumber
k_2&<&(-3\,+\,k_1)/5
\eea
since this would require
$3\,+\,5\,k_1\,<\,-3/5\,+\,k_1/5$,
which requires
$24k_1<-18$ or $k_1<-3/4$.
This contradicts $k_1\ge 0$,
so different values of $|k|$ must yield different
values for ${\cal E}_{12/4}$.

Therefore, at this level of perturbation,
the eigenvalues generically have multiplicity 1
when $k=0$ and multiplicity 2 when $k\ge 1$.

\vskip 5mm
Explicitly,
\bea\nonumber\hspace{-9pt}
{\cal E}_{12/4}&=&
-\quad\frac 1{32\,\mu_1}\
\left(\,\frac{a_3}{a_2}\,\right)^2\
\left(\,11\,+\,30\,n\,+\,30\,n^2\,\right)\quad+\quad
\frac{3\,a_4}{8\,a_2\,\mu_1}\
\left(\,1\,+\,2\,n\,+\,2\,n^2\,\right)
\\[3mm]\nonumber
&&-\quad\frac{b_{0,2,1}^2}{8\,b_{0,2,0}\,b_{0,0,2}\,\mu_1}\
\left(\,2^{\phantom{|}}m(k)\,+\,|k|\,+\,1\,\right)^2
\\[3mm]\nonumber
&&+\quad\frac{b_{0,4,0}}{2\,b_{0,2,0}\,\mu_1}\
\left(\,(\,2\,+\,3\,|k|\,+\,k^2\,)\,+\,
6\,(\,|k|\,+\,1\,)\,m(k)\,+\,6\,m(k)^2\,\right)
\\[3mm]\label{E3}
&&+\quad\frac{X_0}{\mu_2\,Y_0^2}\quad
\sqrt{\frac{b_{0,2,0}}{2\,\mu_1}}\quad
(\,2^{\phantom{|}}m(k)\,+\,|k|\,+\,1\,).
\eea

\pagebreak
\vskip 6mm \noindent
{\bf Order ${\bf \eps^{13/4}}$}
\bea\nonumber
&&\hspace{-25pt}[H_{0,x}-{\cal E}_{6/4}]\ \psi_{7/4}\ +\
[H_{0,r,\gamma}-{\cal E}_{8/4}]\ \psi_{5/4}\ +\
[H_{0,y}-{\cal E}_{10/4}]\ \psi_{3/4}
\\[3mm]\nonumber
&&\hspace{-2pt}-\quad
\frac 1{2\,\mu_2\,Y_0}\
\frac{\partial \psi_0}{\partial y}\ +\
a_3\ x^3\ \psi_{4/4}\ +\
a_4\ x^4\ \psi_{1/4}\ +\
b_{1,0,1}\ x\,y\ \psi_{3/4}\ +\
b_{0,2,1}\ r^2\,y\ \psi_{2/4}
\\[3mm]\nonumber
&&\hspace{-2pt}+\quad
b_{1,2,0}\ x\,r^2\ \psi_{2/4}\ +\
b_{0,4,0}\ r^4\ \psi_{1/4}\ +\
b_{0,0,3}\ y^3\,\psi_0\ +\
b_{1,0,2}\ x\,y^2\,\psi_0\ +\
b_{2,0,1}\ x^2\,y\,\psi_0
\\[3mm]\nonumber
&&\hspace{-2pt}+\quad
\frac{X_0^2}{2\,\mu_2\,Y_0^2}\
H_{0,r,\gamma}\ \psi_{1/4}
\ +\
\frac{X_0}{\mu_2\,Y_0^2}\ \left(\
r\ \frac{\partial^2\phantom{i}}{\partial x\,\partial r}\ +\
\frac{\partial\phantom{i}}{\partial x}\ \right)\ \psi_0
\\[3mm]\nonumber
&&\hspace{-25pt}=\quad
{\cal E}_{13/4}\ \psi_0\ +\
{\cal E}_{12/4}\ \psi_{1/4}.
\eea

\vskip 6mm \noindent
{\bf Order ${\bf \eps^{14/4}}$}
\bea\nonumber
&&\hspace{-25pt}[H_{0,x}-{\cal E}_{6/4}]\ \psi_{8/4}\ +\
[H_{0,r,\gamma}-{\cal E}_{8/4}]\ \psi_{6/4}\ +\
[H_{0,y}-{\cal E}_{10/4}]\ \psi_{4/4}
\\[3mm]\nonumber
&&\hspace{-10pt}\quad -\quad
\frac 1{2\,\mu_2\,Y_0}\
\frac{\partial \psi_{1/4}}{\partial y}\ +\
a_3\ x^3\ \psi_{5/4}\ +\
a_4\ x^4\ \psi_{2/4}\ +\
b_{1,0,1}\ x\,y\ \psi_{4/4}\ +\
b_{2,0,1}\ x^2\,y\,\psi_{1/4}
\\[3mm]\nonumber
&&\hspace{-10pt}\quad +\quad
b_{0,2,1}\ r^2\,y\ \psi_{3/4}\ +\
b_{1,2,0}\ x\,r^2\ \psi_{3/4}\ +\
b_{0,4,0}\ r^4\ \psi_{2/4}\ +\
b_{0,0,3}\ y^3\ \psi_{1/4}
\\[3mm]\nonumber
&&\hspace{-10pt}\quad +\quad
b_{1,0,2}\ x\,y^2\ \psi_{1/4}\ +\
b_{0,2,2}\ r^2\,y^2\ \psi_0\ +\
b_{1,2,1}\ x\,r^2\,y\ \psi_0\ +\
b_{2,2,0}\ x^2\,r^2\ \psi_0
\\[3mm]\nonumber
&&\hspace{-10pt}\quad +\quad
\frac{X_0^2}{2\,\mu_2\,Y_0^2}\
H_{0,r,\gamma}\ \psi_{2/4}\ +\
%\\[3mm]\nonumber
%&&\hspace{-10pt}\quad +\quad
\frac{X_0}{\mu_2\,Y_0^2}\ \left(\
r\ \frac{\partial^2\phantom{i}}{\partial x\,\partial r}\ +\
\frac{\partial\phantom{i}}{\partial x}\ \right)\ \psi_{1/4}
\ \,-\ \,\frac{r^2}{2\,\mu_2\,Y_0^2}\
\frac{\partial^2\phantom{i}}{\partial x^2}\ \psi_0
\\[5mm]\nonumber
&&\hspace{-10pt}\quad +\quad
\frac 1{\mu_2\,Y_0^2}\ \Bigg[\ \left(
X_0\,\sin\,\gamma\ \frac{\partial\phantom{i}}{\partial r}\ +\
\frac{X_0}r\,\cos\,\gamma\
\frac{\partial\phantom{i}}{\partial\gamma}\,\right)\left(
\frac 1{\sin\,\theta}\
\frac{\partial\phantom{i}}{\partial\phi}\ -\ \cot\,\theta\
\frac{\partial\phantom{i}}{\partial\gamma}\right)
\\[3mm]\nonumber
&&\hspace{62mm}+\quad\left( X_0\ \cos\,\gamma\
\frac{\partial\phantom{i}}{\partial r}\ -\
\frac{X_0}r\ \sin\,\gamma\
\frac{\partial\phantom{i}}{\partial\gamma}\,\right)
\frac{\partial\phantom{i}}{\partial\theta}\ \Bigg]\ \psi_0
\\[3mm]\nonumber
&&\hspace{-10pt}=\quad
{\cal E}_{14/4}\ \psi_0\ +\
{\cal E}_{13/4}\ \psi_{1/4}\ +\
{\cal E}_{12/4}\ \psi_{2/4}.
\eea

\vskip 1cm\noindent
{\bf Note}:\quad This is where we first encounter operators
that mix the various different values of $k$.
If we use (\ref{Lcoupling}) in the above expression
and take $\psi_0$ to be a linear combination of the two
degenerate states with $|k|=\lambda$, we see that
the last term on the left hand side of the equation contains
$L_{\pm'}\,|\,j,\,j_z,\,\lambda\,\rangle$ and
$L_{\pm'}\,|\,j,\,j_z,\,-\lambda\,\rangle$, which are linear
combinations of $|\,j,\,j_z,\,\lambda\pm 1\,\rangle$ and
$L_{\pm'}\,|\,j,\,j_z,\,-\lambda\pm 1\,\rangle$, respectively.
Thus, $\psi_{6/4}$ is the lowest order term that involves
$k\ne\pm\lambda$.
%When solving for a component of $\psi_{6/4}$,
%we apply the reduced resolvent of
%$\left[L_{z'}^2\,-\,\lambda^2\right]$ for the first time.

\pagebreak

\vskip 6mm \noindent
{\bf Order ${\bf \eps^{15/4}}$}
\bea\nonumber
&&\hspace{-25pt}[H_{0,x}-{\cal E}_{6/4}]\ \psi_{9/4}\ +\
[H_{0,r,\gamma}-{\cal E}_{8/4}]\ \psi_{7/4}\ +\
[H_{0,y}-{\cal E}_{10/4}]\ \psi_{5/4}
\\[3mm]\nonumber
&&\hspace{-18pt}\quad -\quad
\frac 1{2\,\mu_2\,Y_0}\
\frac{\partial \psi_{2/4}}{\partial y}\ +\
a_3\ x^3\ \psi_{6/4}\ +\
a_4\ x^4\ \psi_{3/4}\ +\
a_5\ x^5\ \psi_0\ +\
b_{1,0,1}\ x\,y\ \psi_{5/4}
\\[3mm]\nonumber
&&\hspace{-18pt}\quad +\quad
b_{0,2,1}\ r^2\,y\ \psi_{4/4}\ +\
b_{1,2,0}\ x\,r^2\ \psi_{4/4}\ +\
b_{0,4,0}\ r^4\ \psi_{3/4}\ +\
b_{0,0,3}\ y^3\ \psi_{2/4}
\\[3mm]\nonumber
&&\hspace{-18pt}\quad +\quad
b_{2,0,1}\ x^2\,y\,\psi_{2/4}\ +\
b_{1,0,2}\ x\,y^2\ \psi_{2/4}\ +\
b_{0,2,2}\ r^2\,y^2\ \psi_{1/4}\ +\
b_{1,2,1}\ x\,r^2\,y\ \psi_{1/4}
\\[3mm]\nonumber
&&\hspace{-18pt}\quad +\quad
b_{2,2,0}\ x^2\,r^2\ \psi_{1/4}\ +\
b_{0,4,1}\ r^4\,y\ \psi_0\ +\
b_{1,4,0}\ x\,r^4\ \psi_0
\\[3mm]\nonumber
&&\hspace{-18pt}\quad +\quad
\frac{X_0^2}{2\,\mu_2\,Y_0^2}\
H_{0,r,\gamma}\ \psi_{3/4}\ +\
%\\[3mm]\nonumber
%&&\hspace{-18pt}\quad +\quad
\frac{X_0}{\mu_2\,Y_0^2}\ \left(\
r\ \frac{\partial^2\phantom{i}}{\partial x\,\partial r}\ +\
\frac{\partial\phantom{i}}{\partial x}\ \right)\ \psi_{2/4}
\ \,-\ \,\frac{r^2}{2\,\mu_2\,Y_0^2}\
\frac{\partial^2\phantom{i}}{\partial x^2}\ \psi_{1/4}
\\[5mm]\nonumber
&&\hspace{-18pt}\quad +\quad
\frac 1{\mu_2\,Y_0^2}\ \Bigg[\ \left(
X_0\,\sin\,\gamma\ \frac{\partial\phantom{i}}{\partial r}\ +\
\frac{X_0}r\,\cos\,\gamma\
\frac{\partial\phantom{i}}{\partial\gamma}\,\right)\left(
\frac 1{\sin\,\theta}\
\frac{\partial\phantom{i}}{\partial\phi}\ -\ \cot\,\theta\
\frac{\partial\phantom{i}}{\partial\gamma}\right)
\\[3mm]\nonumber
&&\hspace{59mm}+\quad\left( X_0\ \cos\,\gamma\
\frac{\partial\phantom{i}}{\partial r}\ -\
\frac{X_0}r\ \sin\,\gamma\
\frac{\partial\phantom{i}}{\partial\gamma}\,\right)
\frac{\partial\phantom{i}}{\partial\theta}\ \Bigg]\ \psi_{1/4}
\\[3mm]\nonumber
&&\hspace{-18pt}\quad +\quad\frac{X_0}{\mu_2\,Y_0^2}\ \left(\
x\ -\ \frac{X_0\,y}{Y_0}\ \right)\ \left(\
-\ \frac{\partial^2\phantom{i}}{\partial r^2}
\ -\ \frac 1r\
\frac{\partial\phantom{i}}{\partial r}
\ +\ \frac 1{r^2}\ L^2_{z'}\ \right)\ \psi_0
\\[5mm]\nonumber
&&\hspace{-18pt}\quad +\quad\frac 1{\mu_2\,Y_0^2}\ \Bigg[\ \left(
-\ r\ \sin\,\gamma\ \frac{\partial\phantom{i}}{\partial x}\
\right)\ \left(\ \frac 1{\sin\,\theta}\
\frac{\partial\phantom{i}}{\partial\phi}\ -\ \cot\,\theta\
\frac{\partial\phantom{i}}{\partial\gamma}\ \right)
\\[3mm]\nonumber
&&\hspace{105mm}\ -\ r\ \cos\,\gamma\ \
\frac{\partial^2\phantom{i}}{\partial x\,\partial\theta}\
\Bigg]\ \psi_0
\\[4mm]\nonumber
&&\hspace{-25pt}=\quad
{\cal E}_{15/4}\ \psi_0\ +\
{\cal E}_{14/4}\ \psi_{1/4}\ +\
{\cal E}_{13/4}\ \psi_{2/4}\ +\
{\cal E}_{12/4}\ \psi_{3/4}.
\eea

\pagebreak

\vskip 6mm \noindent
{\bf Order ${\bf \eps^{16/4}}$}
\bea\nonumber
&&\hspace{-29pt}[H_{0,x}-{\cal E}_{6/4}]\ \psi_{10/4}\ +\
[H_{0,r,\gamma}-{\cal E}_{8/4}]\ \psi_{8/4}\ +\
[H_{0,y}-{\cal E}_{10/4}]\ \psi_{6/4}
\\[3mm]\nonumber
&&\hspace{-3pt}\qquad -\quad
\frac 1{2\,\mu_2\,Y_0}\
\frac{\partial \psi_{3/4}}{\partial y}\ +\
a_3\ x^3\ \psi_{7/4}\ +\
a_4\ x^4\ \psi_{4/4}\ +\
a_5\ x^5\ \psi_{1/4}\ +\
b_{1,0,1}\ x\,y\ \psi_{6/4}
\\[3mm]\nonumber
&&\hspace{-3pt}\qquad +\quad
b_{0,2,1}\ r^2\,y\ \psi_{5/4}\ +\
b_{1,2,0}\ x\,r^2\ \psi_{5/4}\ +\
b_{0,4,0}\ r^4\ \psi_{4/4}\ +\
b_{2,0,1}\ x^2\,y\ \psi_{3/4}
\\[3mm]\nonumber
&&\hspace{-3pt}\qquad +\quad
b_{0,0,3}\ y^3\ \psi_{3/4}\ +\
b_{1,0,2}\ x\,y^2\ \psi_{3/4}\ +\
b_{0,2,2}\ r^2\,y^2\ \psi_{2/4}\ +\
b_{1,2,1}\ x\,r^2\,y\ \psi_{2/4}
\\[3mm]\nonumber
&&\hspace{-3pt}\qquad +\quad
b_{2,2,0}\ x^2\,r^2\ \psi_{2/4}\ +\
b_{0,4,1}\ r^4\,y\ \psi_{1/4}\ +\
b_{1,4,0}\ x\,r^4\ \psi_{1/4}\ +\
b_{0,6,0}\ r^6\ \psi_0
\\[3mm]\nonumber
&&\hspace{-3pt}\qquad +\quad
b_{0,0,4}\ y^4\ \psi_0\ +\
b_{1,0,3}\ x\,y^3\ \psi_0\ +\
b_{2,0,2}\ x^2\,y^2\ \psi_0\ +\
b_{3,0,1}\ x^3\,y\ \psi_0
\\[3mm]\nonumber
&&\hspace{-3pt}\qquad +\quad
\frac{X_0^2}{2\,\mu_2\,Y_0^2}\ H_{0,r,\gamma}\ \psi_{4/4}
%\\[3mm]\nonumber
%&&\hspace{-3pt}\qquad +\quad
\ +\ \frac{X_0}{\mu_2\,Y_0^2}\,\left(
r\ \frac{\partial^2\phantom{i}}{\partial x\,\partial r}\,+\,
\frac{\partial\phantom{i}}{\partial x}\right)\,\psi_{3/4}
\ -\ \frac{r^2}{2\,\mu_2\,Y_0^2}\
\frac{\partial^2\phantom{i}}{\partial x^2}\ \psi_{2/4}
\\[3mm]\nonumber
&&\hspace{-3pt}\qquad +\quad
\frac 1{\mu_2\,Y_0^2}\ \Bigg[\ \left(
X_0\,\sin\,\gamma\ \frac{\partial\phantom{i}}{\partial r}\ +\
\frac{X_0}r\,\cos\,\gamma\
\frac{\partial\phantom{i}}{\partial\gamma}\,\right)\left(
\frac 1{\sin\,\theta}\
\frac{\partial\phantom{i}}{\partial\phi}\ -\ \cot\,\theta\
\frac{\partial\phantom{i}}{\partial\gamma}\right)
\\[3mm]\nonumber
&&\hspace{59mm}+\quad\left( X_0\ \cos\,\gamma\
\frac{\partial\phantom{i}}{\partial r}\ -\
\frac{X_0}r\ \sin\,\gamma\
\frac{\partial\phantom{i}}{\partial\gamma}\,\right)
\frac{\partial\phantom{i}}{\partial\theta}\ \Bigg]\ \psi_{2/4}
\\[3mm]\nonumber
&&\hspace{-3pt}\qquad +\quad\frac{X_0}{\mu_2\,Y_0^2}\ \left(\
x\ -\ \frac{X_0\,y}{Y_0}\ \right)\ H_{0,r,\gamma}\ \psi_{1/4}
\\[3mm]\nonumber
&&\hspace{-3pt}\qquad +\quad\frac 1{\mu_2\,Y_0^2}\ \Bigg[\ \left(
-\ r\ \sin\,\gamma\ \frac{\partial\phantom{i}}{\partial x}\
\right)\ \left(\ \frac 1{\sin\,\theta}\
\frac{\partial\phantom{i}}{\partial\phi}\ -\ \cot\,\theta\
\frac{\partial\phantom{i}}{\partial\gamma}\ \right)
\\[3mm]\nonumber
&&\hspace{102mm}\ -\ r\ \cos\,\gamma\ \
\frac{\partial^2\phantom{i}}{\partial x\,\partial\theta}\
\Bigg]\ \psi_{1/4}
\\[2mm]\nonumber
&&\hspace{-3pt}\qquad +\quad\Bigg[\ -\
\frac{2\,X_0\,y}{\mu_2\,Y_0^3}\ \left(\
\frac{\partial\phantom{i}}{\partial x}\ +\ r\
\frac{\partial^2\phantom{i}}{\partial x\,\partial r}\ \right)
\\[3mm]\nonumber
&&\hspace{3cm}+\quad
\frac 1{2\,\mu_2\,Y_0^2}\ \left(\ 2\,x\,r\
\frac{\partial^2\phantom{i}}{\partial x\,\partial r}\ +\
r\ \frac{\partial\phantom{i}}{\partial r}\ +\ 2\,x\
\frac{\partial\phantom{i}}{\partial x}\ -\
L_{z'}^2\ \right)\ \Bigg]\ \psi_0
\\[3mm]\nonumber
&&\hspace{-3pt}\qquad +\quad
\frac{j(j+1)}{2\,\mu_2\,Y_0^2}\
\psi_0\quad+\quad
\frac 1{\mu_2\,Y_0^2}\ y\
\frac{\partial \psi_0}{\partial y}
\quad+\quad D(0)\ \psi_0
\\[6mm]\nonumber
&&=\quad
{\cal E}_{16/4}\ \psi_0\ +\
{\cal E}_{15/4}\ \psi_{1/4}\ +\
{\cal E}_{14/4}\ \psi_{2/4}\ +\
{\cal E}_{13/4}\ \psi_{3/4}\ +\
{\cal E}_{12/4}\ \psi_{4/4}.
\eea

%%%%%%%%%%%%%%%%%%%%%%%%%%%%%%%%%%%%%%%%%%%%%%%%%%%%%%%%%
%%%%%%%%%%%%%%%%%%%%%%%%%%%%%%%%%%%%%%%%%%%%%%%%%%%%%%%%%
\subsection{The Complete Asymptotic Expansion}
%%%%%%%%%%%%%%%%%%%%%%%%%%%%%%%%%%%%%%%%%%%%%%%%%%%%%%%%%
%%%%%%%%%%%%%%%%%%%%%%%%%%%%%%%%%%%%%%%%%%%%%%%%%%%%%%%%%

We now prove
the existence of a complete
expansion in powers of $\eps^{1/4}$ for the quasienergies
and the corresponding quasimodes under suitable hypotheses.
The following proposition completes the proof of Theorem
\ref{semiclassical}.

\begin{prop}\label{formal}
We assume the potential energy surface (\ref{surface})
is smooth, with Taylor series
given by (\ref{t1}) and (\ref{t2}).
Then, the eigenvalue problem for (\ref{bigdog}) can be solved
by formal asymptotic expansions of the form
\bea\nonumber
{\cal E}\ &=&\
\sum_{l=0}^{N}\ \eps^{l/4}\ {\cal E}_{l/4}\ \,+\ \, 
O(\eps^{(N+1)/4}),
\\[1mm] \nonumber
\psi(x,\,r,\,y,\,\theta,\,\phi,\,\gamma)\ &=&\
\sum_{l=0}^{N}\ \eps^{l/4}\
\psi_{l/4}(x,\,r,\,y,\,\theta,\,\phi,\,\gamma)\ \,+\ \,
O(\eps^{(N+1)/4}),
\eea
for any $N\in{\mathbb N}$.
\end{prop}

\noindent
{\bf Proof}\\
Keeping the original variables $(X,R,Y)$, we first make
use of the invariant subspace ${\cal L}$ generated by
the basis $\{| k \ket \}_{k=-j,\cdots, j}$ of eigenvectors
of $L_{z'}$, where we have dropped the fixed parameters
$j$ and $j_z$ from the notation. In this basis, the operator
$J^2-2L\cdot J +L^2$ can be represented by a matrix.
Let ${\mathbb I}$ denote the
identity matrix,  ${A}$ denote the
matrix representation of
$i\,\frac{\sin(\gamma)}{\sin(\theta)}
\left(-i\,\frac{\partial\phantom{|}}{\partial\phi}
+i\,\cos(\theta)
\frac{\partial\phantom{|}}{\partial\gamma}\right)
+\cos(\gamma)
\frac{\partial\phantom{|}}{\partial\theta}$,
and ${B}$ denote the
matrix representation of
$i\,\frac{\cos(\gamma)}{\sin(\theta)}
\left(-i\,\frac{\partial\phantom{|}}{\partial\phi}
+i\,\cos(\theta)
\frac{\partial\phantom{|}}{\partial\gamma}\right)
+\sin(\gamma)
\frac{\partial\phantom{|}}{\partial\theta}$.
Note that these angular differential operators can be
written as linear combinations of $L'_+$ and $L'_-$, which ensures
that they leave ${\cal L}$ invariant.\\

With these definitions, we can write
\bea\nonumber
&&J^2-2L\cdot J +L^2
\\[3mm] \nonumber
&=&\left(j(j+1)+
\left(-R^2\frac{\partial^2\phantom{|}}{\partial X^2}
+2XR\frac{\partial^2\phantom{|}}{\partial RX}-
X^2\frac{\partial^2\phantom{|}}{\partial
R^2}+\left(R-\frac{X^2}{R}\right)\frac{\partial\phantom{|}}{\partial
R}+2X\frac{\partial\phantom{|}}{\partial X}\right)\right){\mathbb I}
\\[3mm] \nonumber
&&+\left(\frac{X^2}{R^2}-1\right)L^2_{z'}-
2\left(R\frac{\partial\phantom{|}}{\partial X}-X
\frac{\partial\phantom{|}}{\partial R}\right){A}
-2\frac{X}{R}{B}.
\eea

Then, going to the rescaled variables and dropping the symbol
$\mathbb I$, the differential operator (\ref{bigdog}) takes the form
\bea\nonumber
&&\hspace{-1cm}-\ \frac{\eps^{6/4}}{2\,\mu_1(\eps)}\
\frac{\partial^2\phantom{|}}{\partial x^2}\ -\
\frac{\eps^{8/4} }{2\,\mu_1(\eps)}\ \left(\,
\frac{\partial^2\phantom{|}}{\partial r^2}\ +\
\frac 1r\ \frac{\partial\phantom{|}}{\partial r}\ -\
\frac 1{r^2}\ L^2_{z'}
\,\right)\ -\
\frac{\eps^{10/4}}{2\,\mu_2(\eps)}\
\frac{\partial^2\phantom{|}}{\partial y^2}
\\[3mm]\nonumber
&&-\
\frac{\eps^{13/4}}{\mu_2(\eps)\,(Y_0+\eps^{3/4}y)}\
\frac{\partial\phantom{|}}{\partial y}\
-\ \frac{\eps^{12/4}(X_0+\eps^{3/4}x)^2}
{2\ \mu_2(\eps)\,(Y_0+\eps^{3/4}y)^2}\
\left\{\frac{\partial^2\phantom{|}}{\partial r^2} +
\frac{1}{r}\frac{\partial\phantom{|}}{\partial r}
-\frac 1{r^2}\
L^2_{z'}  \right\}\\[3mm]\nonumber
&&
+\
\frac{\eps^{13/4}(X_0+\eps^{3/4}x)}
{\mu_2(\eps)\,(Y_0+\eps^{3/4}y)^2}\
\left\{r
\frac{\partial^2\phantom{|}}{\partial x\partial r}
+\frac{\partial\phantom{|}}{\partial x}  \right\}
\\[3mm]\nonumber
&&+\
\frac{\eps^{14/4}(X_0+\eps^{3/4}x)}
{\mu_2(\eps)\,(Y_0+\eps^{3/4}y)^2}\
\left\{\frac{\partial\phantom{|}}
{\partial r}\,A\,-\frac{1}{r}\,B\right\}
-\
\frac{\eps^{14/4}}{2\ \mu_2(\eps)\,(Y_0+\eps^{3/4}y)^2}\
\left\{r^2\frac{\partial^2\phantom{|}}
{\partial x^2}\right\}
\\[3mm]\nonumber
&&-\
\frac{\eps^{15/4}}{\mu_2(\eps)\,(Y_0+\eps^{3/4}y)^2}\
\left\{r\frac{\partial\phantom{|}}{\partial x}\,A\,\right\}+\
\frac{\eps^{16/4}}{2\,\mu_2(\eps)\,(Y_0+\eps^{3/4}y)^2}\
\left\{\,j(j+1)+r
\frac{\partial\phantom{|}}{\partial r} - L^2_{z'}\,\right\}
\\[4mm]\nonumber
&&\qquad\qquad\qquad\quad +\quad a_0\
+\ \sum_{j=2}^{\infty}\ a_j\ \eps^{3j/4}\ x^j\ +
\sum_{\scriptsize
\begin{array}{c}j+k+l\ge 2\\ k+l\ge 1\\
k\ \mbox{even}\end{array}}\,
b_{j,\,k,\,l}\ \eps^{1+\frac{3(j+l)+2k}4}\ x^j\,r^k\,y^l.
\eea

%{\bf Alain's comment:}\quad{\it
%I think we should substantiate this paragraph. It 
%should be enough to point out that the functions of
%``$r$" we'll have to  integrate will be integrated
%with the measure ``$r\,dr$", are given by 
%polynomials in ``$r$" times $exp(-r^2/2)$,
%or here first and second derivatives, 
%so that the only potentially dangerous term is
%$L_z / r^2$. But it is non trivial only when it acts
%on $\Phi(|k|,r)$ with $|k|>0$, so that there is no 
%problem.}

We get a matrix valued differential operator given as
a formal infinite series in powers of $\eps^{1/4}$ by
expanding the reduced masses $\mu_j(\eps)$ and
the denominators $(Y_0+\eps^{3/4})$ and $(Y_0+\eps^{3/4})^2$.
Observe that in each
term of the resulting expansion, the differential operators
are at most of order two.

The $r$ dependence of these operators is explicit,
which will allow us to
check that that the factors $1/r$ and $1/r^2$ do not cause
divergences in the expressions that we encounter below.
The measure in the $r$ variable is $r\,dr$, so the only
term that might yield a vector not in $L^2$ is the
$L_{z'}/r^2$. In the eigenspace where $L_{z'}$ multiplies
by zero, there is no problem. In the eigenspaces where
$L_{z'}$ multiplies by something non-zero, the wave functions
contain factors of $r$, so again, there is no problem.

We introduce the notation
\bea\nonumber
\Psi_{l/4}(x,r,y)&=&\sum_{k=-j}^j\,
\psi_{l/4}(x,r,y,k)\,|k\ket\ \equiv\
\pmatrix{\psi_{l/4}(x,r,y,-j)\vspace{1mm}\cr
\psi_{l/4}(x,r,y,-j+1)\vspace{1mm}\cr
\vdots\vspace{1mm}\cr
\psi_{l/4}(x,r,y,j)}.
\eea

We have already explicitly presented perturbation theory through
order $\eps^{l/4}$ for $l\le 11$.
The equation we must solve at order $\eps^{l/4}$ with $l\ge 12$
now can be expressed as
\bea\nonumber
&&(H_{0,x}-{\cal E}_{6/4})\,\Psi_{(l-6)/4}\ +\ 
(H_{0,r,\gamma}-{\cal E}_{8/4})\,\Psi_{(l-8)/4}\ +\
(H_{0,y}-{\cal E}_{10/4})\,\Psi_{(l-10)/4}
\\[2mm] \nonumber
&&+\,a_3\, x^3\,\Psi_{(l-9)/4}\ +\ b_{1,0,1}\,x\,y\,\Psi_{(l-10)/4}
\ +\ \sum_{q=11}^l\,D_{q}\,\Psi_{(l-q)/4}
\\[3mm] \label{geq}
&=&%\hspace{5cm}
{\cal E}_{l/4}\,\Psi_{0/4}\ +\ \cdots\ +\
{\cal E}_{12/4}\,\Psi_{(l-12)/4},
\eea
where the symbols $D_q$ denote at most second order 
differential operators in $x,r,y$ with matrix valued coefficients
whose entries are polynomials in these
variables divided by $r^p$, with $p=0, 1, 2$.
We note also that  $H_{0,r,\gamma}$ is now matrix--valued,
because of the centrifugal term $L^2_{z'}/r^2$,
whereas $H_{0,x}$ and $H_{0,y}$ are scalar
differential operators multiplied by the identity matrix.

The point of this decomposition is to separate the vectors
$\Psi_{q/4}$ of order less than or equal to $(l-11)/4$ from those
of order $(l-10)/4$ to $(l-6)/4$.

Let $P_x$, $P_y$ and $P_{r,\gamma}$ be the
orthogonal projectors on the eigenstates
$\Phi_1(x)$, $\Phi_3(y)$ and on the subspace
$Z_0=\mbox{span}\{\Phi_2(r,|k|)\ |k\ket\}_{k\in K}$, respectively.
We abuse notation and use the same symbols to
denote the corresponding projectors when considered on
$L^2({\mathbb R}_x, dx)\otimes
L^2({\mathbb R}^+_r, rdr)\otimes
L^2({\mathbb R}_y, dy)\otimes {\cal L}$.
Note that these operators
commute with one another and that the following
identity holds for any $q\in\mathbb N$\,:
\be\label{pxp}
P_x\,x^{2q+1}\,=\,P_x\,x^{2q+1}\,P_x^{\perp},
\qquad\mbox{where }\quad P_x^{\perp}={\mathbb I}-P_x.
\ee
Also, we have constructed $\Psi_{l/4}$ so that
\be\label{ppp}
\Psi_0\ =\ P_x\,P_{r,\gamma}\,P_y\,\Psi_0\quad\mbox{and}\quad
P_x\,P_{r,\gamma}\,P_y\,\Psi_{l/4}\ =\ 0,\qquad\mbox{for all}\quad
l\geq 1.
\ee
Hence, for $l\geq 1$,
\be\label{decp}
\Psi_{l/4}\ =\ P_x^\perp\,
\Psi_{l/4}\,+\,P_x\,P_{r,\gamma}^\perp\,\Psi_{l/4}\,+\,
P_x\,P_{r,\gamma}\,P_y^\perp\,\Psi_{l/4}.
\ee
In terms of the quantities introduced in the explicit
computations of the lower orders, we have in particular
\bea\label{paral}
P_x\,\Psi_{l/4}&=&\sum_{k=-l}^l\
\Phi_1(x)\ f_{l/4}(r,y,k)\ |k\ket
\\[2mm]\nonumber
P_x\,P_{r,\gamma}\,\Psi_{l/4}
&=&\sum_{k\in K}\ \Phi_1(x)\
\Phi_2(r,|k|)\ g_{l/4}(y,k)\ | k\ket
\\[2mm]\nonumber
P_x\,P_{r,\gamma}\,P_y\,\Psi_{0}
&=&\sum_{k\in K}\ \Phi_1(x)\ \Phi_2(r,|k|)\
\Phi_3(y)\,c_{k,0}\ | k\ket,
\eea
where $c_{k,0}\in{\mathbb C}$ and
$\sum_{k\in K} |c_{k,0}|^2=1.$
Note that by virtue of (\ref{ppp}),
\be\label{gperp}
g_{l/4}(y,k)\ =\ P^\perp_y\,g_{l/4}(y,k),
\quad\mbox{for any $k\in K$ and  any $l>0$}.
\ee

We solve (\ref{geq}) by two independent steps.
The first consists
of determining the vectors $\Psi_{l/4}$ for any
set of coefficients $\{c_{0,k}\}_{k\in K}$,
and the other consists of solving an eigenvalue  
equation for ${\cal E}_{j/4}$ in $\C^{\#(K)}$ which may
reduce the set of free coeffcients $\{c_{0,k}\}_{k\in K}$.
It is only when we construct the actual  
quasimode that we restrict the values of the
coefficients $\{c_{0,k}\}_{k\in K}$ to those given by the
determination of the the ${\cal E}_{l/4}$'s.

We now formulate our induction hypothesis for $l\ge 12$.

\noindent
{\bf IH:}\quad
After solving  equation (\ref{geq}) through order
$\eps^{(l-1)/4}$ for vectors satisfying
(\ref{ppp}), we have:
\begin{itemize}
\item The following vectors
are determined completely in terms of the coefficients
$\{c_{0,k}\}_{k\in K}$
and depend linearly on $\{c_{0,k}\}_{k\in K}$ :
\be\nonumber
\begin{array}{ll}\nonumber
\Psi_{q/4}, &\quad\mbox{for}\quad q=0,\,1,\,\cdots,\,l-11,
\\[1mm] \nonumber
({\mathbb I}-P_xP_{r,\gamma})\,\Psi_{(l-10)/4},&
\\[1mm] \nonumber
({\mathbb I}-P_x\,P_{r,\gamma})\,\Psi_{(l-9)/4},&
\\[1mm] \nonumber
({\mathbb I}- P_x)\,\Psi_{(l-8)/4},&\quad\mbox{and}
\\[1mm] \nonumber
({\mathbb I}- P_x - P^\perp_x\,P_{r,\gamma})\,\Psi_{(l-7)/4}.&
\end{array}
\ee
\item The $x$ dependence of the vector
$P^\perp_x\,P_{r,\gamma}\,\Psi_{(l-7)/4}$
is determined and has the form
\be
P^\perp_x\,P_{r,\gamma}\,\Psi_{(l-7)/4}\ =\
P^\perp_x\,P_{r,\gamma}\,\Psi_{(l-7)/4}(\{g_{(l-10)/4}\}),
\ee
with linear dependence on $\{g_{(l-10)/4}(y,k)\}_{k\in K}$,
the set of   functions
$\{g_{(l-10)/4}\}$ entailing the unknown $y$ dependence.
\item There exist vector spaces
$W_q\subseteq \C^{\#(K)}$ satisfying
\be\nonumber
\C^{\#(K)}=W_0\supseteq W_1\supseteq\cdots\supseteq W_{l-1}
\ee
such that ${\cal E}_{q/4}$
is determined by an eigenvalue equation in
$W_q$, for $q=0, 1, \cdots, l-1$.
\end{itemize}

\vspace{.5cm}
Our explicit computations show that these properties
are satisfied for $l=12$,
with $W_q=\C^{\#(K)}$, for $q=0,\cdots, 11$.
We now show that the induction hypothesis  holds
at order $\eps^{l/4}$.

Using (\ref{pxp}) and  (\ref{ppp}) and applying
$P_x\,P_{r,\gamma}P_y$ to equation
(\ref{geq}) yields
\bea\nonumber
{\cal E}_{l/4}\,\Psi_0&=&P_x\,P_{r,\gamma}\,P_y
\left(a_3 x^3 P_x^\perp\Psi_{(l-9)/4}+
b_{1,0,1}xy P_x^\perp \Psi_{(l-10)/4}+
\sum_{q=11}^l\,D_q\Psi_{(l-q)/4}  \right).
\eea
We note that for $s=9,\,10$, the vectors
$
P_x^\perp\,\Psi_{(l-s)/4}=
P_x^\perp\,({\mathbb I}-P_xP_{r,\gamma})\,\Psi_{(l-s)/4}
$
are completely determined by {\bf IH}. By {\bf IH} again,
the right hand side depends linearly on the set
$\{c_{0,k}\}_{k\in K}$. Expressing the equation in the basis
$\{\Phi_1(x)\Phi_2(|k|, r)\Phi_3(y)\}_{k\in K}$ of $Z_2$,
we get a finite dimensional eigenvalue equation.
Restricting attention to the subspace $W_{l-1}\subseteq
{\mathbb C}^{\#(K)}$ of free coefficients, we get an eigenvalue
equation in $W_{l-1}$ which we solve to yield
${\cal E}_{l/4}$ and the subspace $W_l\subseteq W_{l-1}$ of free
coefficients.

We now turn to the computation of the vectors.
Application of $P_x\,P_{r,\gamma}\,P_y^\perp$
to equation (\ref{geq}) yields
\bea
&&\hspace{-1cm}P_x\,P_{r,\gamma}\,P_y^\perp\,\Psi_{(l-10)/4}
\\[3mm]\nonumber
\hspace{-5mm}
&\hspace{-24mm}=&
\hspace{-1cm}-\quad(H_{0,y}-{\cal E}_{10/4})^{-1}_r
P_xP_{r,\gamma}P_y^\perp\left(
a_3 x^3 P_x^\perp\Psi_{(l-9)/4}+
b_{1,0,1}xy P_x^\perp \Psi_{(l-10)/4}+
\sum_{q=11}^l\tilde D_q\Psi_{(l-q)/4}  \right)
\eea
where $\tilde D_q = D_q-{\cal E}_{q/4}$.
The right hand side is known by {\bf IH}, and since   
$P_x\,P_{r,\gamma}\,P_y^\perp\,\Psi_{(l-10)/4}=
P_x\,P_{r,\gamma}\,\Psi_{(l-10)/4}$,~
(see (\ref{paral}), (\ref{gperp})),
(\ref{decp}) implies  that $\Psi_{(l-10)/4}$ is
fully determined up to the coefficients
$\{c_{0,k}\}_{k\in K}$. Since the dependence
of $P_x\,P_{r,\gamma}\,\Psi_{(l-10)/4}$ is linear in
the previously determined quantities,
we get by {\bf IH} that $\Psi_{(l-10)/4}$ depends
linearly in the coefficients $\{c_{0,k}\}_{k\in K}$.
Hence, the vector
$P^\perp_x\,P_{r,\gamma}\,\Psi_{(l-7)/4}\,(\{g_{(l-10)/4}\})$
in {\bf IH} is, in turn,  
fully determined, and it depends linearly on the
$\{c_{0,k}\}_{k\in K}$'s. Thus, the same is true
for $({\mathbb I}-P_x)\,\Psi_{(l-7)/4}$.

Application of  $P_x\,P_{r,\gamma}^\perp$
to equation (\ref{geq}) yields
\bea
&&P_x\,P_{r,\gamma}^\perp\,\Psi_{(l-8)/4}\ =\
-\ (H_{0,r,\gamma}-{\cal E}_{8/4})^{-1}_r\,
P_x\,P_{r,\gamma}^\perp\quad
\times
\\ \nonumber
&&\left( (H_{0,y}-{\cal E}_{10/4}) \Psi_{(l-8)/4}
+ a_3 x^3 P_x^\perp\Psi_{(l-9)/4}+
b_{1,0,1}xy P_x^\perp \Psi_{(l-10)/4}+\sum_{q=11}^l\tilde
D_q\Psi_{l-q} \right),
\eea
where, by the same arguments,  the right hand side is
fully determined up to the coefficients
$\{c_{0,k}\}_{k\in K}$, on which it depends
linearly. Now, from {\bf IH} and the identity
$$
P_x\,\Psi_{(l-8)/4}\ =\ P_x\,P_{r,\gamma}\,\Psi_{(l-8)/4}\ +\
P_x\,P_{r,\gamma}^\perp\,\Psi_{(l-8)/4}
$$
we see that $({\mathbb I}-P_xP_{r,\gamma})\,\Psi_{(l-8)/4}$
is fully determined and depends linearly on the
coefficients $\{c_{0,k}\}_{k\in K}$.

Finally, application  of  $P_x^\perp$
to equation (\ref{geq}) yields
\bea
&&P_x^\perp\Psi_{(l-6)/4}\ =\
-\,(H_{0,x}-{\cal E}_{6/4})^{-1}_r\,P_x^\perp\,
\left( \hspace{-.7cm}\phantom{\sum_p}
(H_{0,r,\gamma}-{\cal E}_{8/4})\,
P^\perp_x\,\Psi_{(l-8)/4} +\right.
\\ \nonumber
&&\left.
(H_{0,y}-{\cal E}_{10/4})\,P^\perp_x\,\Psi_{(l-10)/4}\,+\,
a_3 x^3 \Psi_{(l-9)/4}\,+\,
b_{1,0,1}xy  \Psi_{(l-10)/4}\,+\,\sum_{q=11}^l\tilde
D_q\,\Psi_{l-q} \right),
\eea
where,
%we inserted a harmless projector $P_{r,\gamma}^\perp$ in front
%of $\Psi_{(j-8)/4}$ and where
this time, the right hand side is not fully
determined since there is
no projector $P^\perp_x$ acting on $\Psi_{(l-9)/4}$.
However, at this step, $\Psi_{(l-10)/4}$ and
$P^\perp_x\Psi_{(l-8)/4}=
P^\perp_x({\mathbb I}-P_xP_{r,\gamma})\Psi_{(l-8)/4}$
are fully determined and linear in
the $\{c_{0,k}\}_{k\in K}$, so that from
{\bf IH} we see that the only undetermined part comes from
$$
P_x\,P_{r,\gamma}\,\Psi_{(l-9)/4}\ =\
\sum_{k\in K}\,\Phi_1(x)\,\Phi_2(r,|k|)\,
g_{(l-9)/4}(y,k)\,c_{k,0}\ | k\ket.
$$

We conclude that the $x$ dependence of the vector
$P_x^\perp\,\Psi_{(l-6)/4}$ is determined,
and that the undetermined part of this
vector depends on the set of functions
$\{g_{(l-9)/4}(y,k)\}_{k\in K}$  
purely linearly.

Thus, we have reproduced the all the requirements
of the induction hypothesis, which ends the proof. \ep

\vspace{.5cm}

%%%%%%%%%%%%%%%%%%%%%%%%%%%%%%%%%%%%%%%%%%%%%%%%%%%%%%%%%
%%%%%%%%%%%%%%%%%%%%%%%%%%%%%%%%%%%%%%%%%%%%%%%%%%%%%%%%%
\subsection{The Expansion Around a Local Minimum}
%%%%%%%%%%%%%%%%%%%%%%%%%%%%%%%%%%%%%%%%%%%%%%%%%%%%%%%%%
%%%%%%%%%%%%%%%%%%%%%%%%%%%%%%%%%%%%%%%%%%%%%%%%%%%%%%%%%

We now describe the construction of quasimodes of
arbitrarily high order under assumptions that are only
local. This construction uses the formal expansions
of Proposition \ref{formal} and the insertion of cutoff functions.
The construction  is quite similar to that given in
\cite{hagjoy10}, so we refrain from presenting all details.

Let $N\geq 0$ be fixed and set 
\bea\label{trunk} \nonumber
\Psi^{(N)}(x,r,y,\theta,\phi,\gamma)
&=&\sum_{l=0}^N\ \eps^{l/4}\ \psi_{l/4}(x,r,y,\theta,\phi,\gamma),
\\
{\cal E}^{(N)}&=&\sum_{l=0}^N\ \eps^{l/4}\ {\cal E}_{l/4},
\\ \nonumber
V^{(N)}(X,Y,R)&=&\hspace{-.3cm}
\sum_{l\leq (N+1)/3}a_l (X-X_0)^l+\eps\hspace{-.9cm}
\sum_{\scriptsize
\begin{array}{c}j+k+l\ge 2\\ k+l\ge 1\\
k\ \mbox{even}\\
4+3(j+l)+2k\leq N
\end{array}}\hspace{-.7cm}
b_{j,\,k,\,l} (X-X_0)^j\,R^k\,(Y-Y_0)^l,
\eea
where the vectors $\psi_{l/4}$ and the scalars ${\cal E}_{l/4}$
are defined in Proposition \ref{formal}.

Then we introduce a cutoff function. Let
${\cal F}:\R\rightarrow [0,1]$ be  $C^\infty$ and 
such that $\mbox{supp }{\cal F}\subset [-2,2]$~ with~
${\cal F}(t)=1$~ for~ $t\in [-1,1]$.~ We set
$$
{\cal F}_\eps(X,R,Y)\ =\ {\cal F}((X-X_0)/\eps^{\delta_1})\
{\cal F}(R/\eps^{\delta_2})\ {\cal F}((Y-Y_0)/\eps^{\delta_3}),
$$
where~ $0<\delta_1<3/4$,~ $0<\delta_2<1/2$~ and~ $0<\delta_3<3/4$.

The quasimode $\Psi_Q^{(N)}$ is defined as
\bea\label{quasimodo}
&&\Psi_Q^{(N)}(X,R,Y,\theta, \phi, \gamma)\\ \nonumber
&&\hspace{1cm}=\quad\eps^{-5/4}\ {\cal F}_\eps(X,R,Y)\
\Psi^{(N)}((X-X_0)/\eps^{3/4},R/\eps^{1/2},(Y-Y_0)/\eps^{3/4},
\theta,\phi,\gamma).
\eea
The factor of $\eps^{-5/4}$ in this expression ensures
asymptotic normalization of the quasimode because of the
Jacobian factor in the integral for the $L^2$ norm.

\begin{prop}\label{local} 
Let
$$
H(\eps)\ =\
-\ \frac{\eps^3}{2\,\mu_1(\eps)}\ \Delta_{(X_1,\,X_2,\,X_3)}\
-\ \frac{\eps^4}{2\,\mu_2(\eps)}\ \Delta_{(Y_1,\,Y_2,\,Y_3)}\
+\ V_1(X)\ +\ \eps \ V_2(X,\,R,\,Y),
$$
satisfy the hypotheses of Proposition \ref{formal}.
Then, for any $N\in\N$, there exists a constant $C_N$, such that
the  vector (\ref{quasimodo}) and the scalar (\ref{trunk}) satisfy~
$\|\Psi_Q^{(N)}\|=1+O(\eps^{1/4})$~ and 
$$
\frac{\left\|H(\eps)\Psi_Q^{(N)}-{\cal E}^{(N)}\Psi_Q^{(N)}\right\|}
{\left\|\Psi_Q^{(N)}\right\|}\ \leq\
C_N\ \eps^{(N+1)/4},\qquad\mbox{as~ $\eps\ra 0$.}
$$
\end{prop}

\noindent
{\bf Proof}\\
We begin by computing the norm of $\Psi_Q^{(N)}$.
The vectors $\psi_{l/4}$, for $l=0,\cdots, N$, are given
as a finite linear combinations of angular functions
$|k,j_z,j\ket$,~ ($k=-j, \cdots, j$), multiplied by Gaussians
in $x, r, y$, times polynomials in these variables.
Thus, they all belong to $L^2$. 

In particular, by our choices for $\psi_0$, we have
\bea\nonumber
&&\int \ |\eps^{-5/4}\,\psi_0((X-X_0)/\eps^{3/4},R/\eps^{1/2},
(Y-Y_0)/\eps^{3/4}, \theta, \phi, \gamma)|^2\ R
\,dR\, dX\, dY\, d\Omega
\\[3mm] \nonumber
&=&\int\ |\psi_0(x,r,y,\theta, \phi, \gamma)|^2\
r\, dr\,dx\,dy\, d\Omega
\\[3mm] \nonumber
&=&1,
\eea
where $d\Omega$ denotes the solid angle element in the angular
variables. The norms of the other $\psi_{l/4}$ are similarly $O(1)$. 

Hence~
$
\|\Psi_Q^{(N)}\|^2=\|\Psi^{(N)}+({\cal F}_\eps^2-1)\Psi^{(N)}\|^2,
$~ where,
\bea\label{small}
&&\left\|(1-{\cal F}_\eps^2)\,\Psi^{(N)}\right\|^2
\\[3mm] \nonumber
&\leq&\int_{\scriptsize  \begin{array}{l}\phantom{b}\\
|X-X_0|\geq \eps^{\delta_1}\\ R\geq \eps^{\delta_2}\\
|Y-Y_0|\geq \eps^{\delta_3}\end{array} } \hspace{-0.5cm}
|\Psi^{(N)}((X-X_0)/\eps^{3/4},R/\eps^{1/2},(Y-Y_0)/\eps^{3/4},
\theta, \phi, \gamma)|^2\,R\,dR\,dX\,dY\,d\Omega.
\eea
The choice of exponents $\delta_j$ and the exponential decay
of $\Psi^{(N)}$  imply that (\ref{small}) is of order $\eps^\infty$,
and we finally see that
$$
\left\|\Psi_Q^{(N)}\right\|\ =\ 1\ +\ O(\eps^{1/4}).
$$

By construction, there exist $C>0$ and $D>0$,
independent of $\eps$, such that 
$$
{\cal R}^{(N)}(X,R,Y)\ =\ V_1(X)\,+\,\eps\,V_2(X,R,Y)\,-\,V^{(N)}(X,Y,R)
$$
satisfies 
\be\label{esr}
|{\cal R}^{(N)}(X,R,Y)|\leq C(|X-X_0|^{(N+1)/3}+
\eps |X-X_0|^{a}R^b|Y-Y_0|^c),
\ee 
where $ 4+ 3(a+c)+2c \geq N+1$, if $(|X-X_0|+R+|Y-Y_0|)<D$.
Consider now 
\bea\nonumber
V\,\Psi_Q^{(N)}&=&V^{(N)}\,\Psi_Q^{(N)}\ +\ {\cal R}^{(N)}\,\Psi_Q^{(N)}
\\[2mm]\nonumber
&=&V^{(N)}\,{\cal F}_\eps\,\Psi^{(N)}\ +\
{\cal R}^{(N)}\,{\cal F}_\eps\,\Psi^{(N)} .
\eea
Due to the support conditions imposed by the cutoff,
we can estimate
${\cal F}_\eps{\cal R}^{(N)}$ by means of (\ref{esr}),
and, after passing to the rescaled variables 
$x,r,y$, we obtain
$$
{\cal F}_\eps(X,R,Y)\,|{\cal R}^{(N)}(X,R,Y)|\ \leq\
{\cal F}_\eps(X,R,Y)\,\eps^{(N+1)/4}\,
C\,\left( |x|^{(N+1)/3}+|x|^ar^b|y|^c\right).
$$
Once again using the Gaussian decay of $\Psi^{(N)}$,
we finally get the $L^2$ estimate
$$
\left\|\,{\cal R}^{(N)}\,\Psi_Q^{(N)}\,\right\|\ =\
O\left(\eps^{(N+1)/4}\right).
$$

We now have estimated everything except the terms in which
the kinetic energy acts on the cutoffs.
First note that derivatives with respect to angular
variables do not affect the cutoffs. Next, by the Leibniz formula,
the first and second derivatives with respect to $x,\ y$, or $r$
acting on ${\cal F}_\eps\,\Psi^{(N)}$ yield supplementary terms given
by  first and second derivatives of ${\cal F}_\eps$ multiplied by
$\Psi^{(N)}$ or first derivatives of $\Psi^{(N)}$. By construction
of the cutoff, the successive derivatives of ${\cal F}_\eps$ are
supported away of the origin in at least one of the variables
$x,\ y$, or $r$. Since $\Psi^{(N)}$ and its derivatives are Gaussian
times polynomials in these variables, these supplementary terms are
all of order $\eps^\infty$.

Finally, taking into account the formal expansions of
Theorem \ref{formal}, and the definition
\bea\nonumber
%H(\eps)&=&
%-\ \frac{\eps^3}{2\,\mu_1(\eps)}\ \Delta_{(X_1,\,X_2,\,X_3)}\
%-\ \frac{\eps^4}{2\,\mu_2(\eps)}\ \Delta_{(Y_1,\,Y_2,\,Y_3)}\
%+\ V_1(X)\ +\ \eps \ V_2(X,\,R,\,Y,),\\
H^{(N)}(\eps)&=&
-\ \frac{\eps^3}{2\,\mu_1(\eps)}\ \Delta_{(X_1,\,X_2,\,X_3)}\
-\ \frac{\eps^4}{2\,\mu_2(\eps)}\ \Delta_{(Y_1,\,Y_2,\,Y_3)}\
+\ V^{(N)}(X,\,R,\,Y,),
\eea
we get the $L^2$ norm estimate
\bea\nonumber
&&\left\| H(\eps)\Psi_Q^{(N)}-{\cal E}^{(N)}\Psi_Q^{(N)}\right\|
\\[2mm]\nonumber
&=&
\left\| H^{(N)}(\eps)\Psi_Q^{(N)}-{\cal E}^{(N)}\Psi_Q^{(N)}\right\|
\ +\ O\left(\eps^{(N+1)/4}\right)
\\[2mm]\nonumber
&=&\left\|{\cal F}_\eps (H^{(N)}(\eps)\Psi^{(N)}-
{\cal E}^{(N)}\Psi^{(N)})\right\|\ +\
O\left(\eps^{(N+1)/4}\right)\ +\ O\left(\eps^\infty\right)
\\[2mm]\nonumber
&=&O\left(\eps^{(N+1)/4}\right).
\eea
\phantom{Q}\hfill\ep
%In the last step, we used again the Gaussian falloff
%of all the vectors involved.\ep

\vskip 1cm
%%%%%%%%%%%%%%%%%%%%%%%%%%%%%%%%%%%%%%%%%%%%%%%%%%%%%%%%%
%%%%%%%%%%%%%%%%%%%%%%%%%%%%%%%%%%%%%%%%%%%%%%%%%%%%%%%%%
\section{Inclusion of the Electrons}\label{Sect4}
\setcounter{equation}{0}
%%%%%%%%%%%%%%%%%%%%%%%%%%%%%%%%%%%%%%%%%%%%%%%%%%%%%%%%%
%%%%%%%%%%%%%%%%%%%%%%%%%%%%%%%%%%%%%%%%%%%%%%%%%%%%%%%%%

In this section we show that including the quantum mechanical
treatment of the electrons does not change the expression
for the energy up to an error of order $\eps^3$.

We decompose the Hamiltonian for all the particles in the
molecule as the sum of the nuclear kinetic energy plus
a self-adjoint electron Hamiltonian
$h_1(Y,\,\theta,\,\phi,\,R,\,\gamma,\,X)$. The electron
Hamiltonian depends parametrically on
$(Y,\,\theta,\,\phi,\,R,\,\gamma,\,X)$ and acts on
functions of all of the electron variables, that we
describe jointly with the single symbol $Z$. To avoid
questions about Berry phases, we assume
$h_1(Y,\,\theta,\,\phi,\,R,\,\gamma,\,X)$ commutes with
complex conjugation, {\it i.e.}, it is a real symmetric
operator.

Because of rotational symmetries, the electron Hamiltonian
can be written as
$$
h_1(Y,\,\theta,\,\phi,\,R,\,\gamma,\,X)\ =\
U(\theta,\,\phi,\,\gamma)\ h_2(X,\,R,\,Y)\
U(\theta,\,\phi,\,\gamma)^{-1},
$$
where $U(\theta,\,\phi,\,\gamma)$ is unitary on
the electron Hilbert space and depends smoothly
on~$\theta$,~$\phi$,~and~$\gamma$. As a consequence,
discrete eigenvalues of
$h_1(Y,\,\theta,\,\phi,\,R,\,\gamma,\,X)$ do not
depend on~$\theta$,~ $\phi$,~ or~ $\gamma$.

We assume that the resolvent of $h_2(X,\,R,\,Y)$
depends smoothly on $(X,\,R,\,Y)$. As a result,
all discrete eigenvalues of
$h_1(Y,\,\theta,\,\phi,\,R,\,\gamma,\,X)$ depend
smoothly on the nuclear configurations.

We assume further that the ground state eigenvalue
$V(X,\,R,\,Y)$ of
$h(Y,\,\theta,\,\phi,\,R,\,\gamma,\,X)$ is discrete
and non-degenerate for
each fixed value of
$(Y,\,\theta,\,\phi,\,R,\,\gamma,\,X)$.
We also assume that $V(X,\,R,\,Y)$ has a
global minimum
at $(X_0,\,0,\,Y_0)$ with a strictly positive
Hessian at that minimum.
To ensure that we are approximating discrete
eigenvalues for the full molecular Hamiltonian,
we assume that the
$V(X_0,\,0,\,Y_0)$ is strictly below the bottom
of the spectrum of $h_2(X,\,R,\,Y)$ for all
$(X,\,R,\,Y)$ outside a small neighborhood of
$(X_0,\,0,\,Y_0)$.

We now introduce $\eps$--dependence in $h_2$, and
hence $h_1$. We choose functions $V_1(X)$ and
$V_2(X,\,R,\,Y)$ that satisfy
$$
V(X,\,R,\,Y)\ =\
V_1(X)\ +\ \eps_0\ V_2(X,\,R,\,Y)
$$
and the restrictions imposed after expression
(\ref{surface}).
Here $\eps_0$ is a fixed value of $\eps$ that
we take to be the fourth root of the electron
mass divided by the carbon $C^{12}$ nuclear mass.
We then define
$h(\eps,\,Y,\,\theta,\,\phi,\,R,\,\gamma,\,X)$
by replacing $V(X,\,R,\,Y)$ by
$V_1(X)\,+\,\eps\ V_2(X,\,R,\,Y)$ in the spectral
decomposition of
$h_1(Y,\,\theta,\,\phi,\,R,\,\gamma,\,X)$. Thus,
we only introduce $\eps$--dependence in this single
eigenvalue and alter none of the eigenfunctions.

\vskip 5mm\noindent
{\bf Remark}\quad
To minimize technicalities, we have made
assumptions for all $(X,\,R,\,Y)$.
At the expense of inserting cut off functions,
our assumptions need only be imposed for
$(X,\,R,\,Y)$ in a neighborhood of $(X_0,\,0,\,Y_0)$.

\vskip 5mm
We shall write down an explicit quasimode
with an $O(\eps^{12/4})$ energy error for
the Schr\"o\-ding\-er operator
$$
H(\eps)\ =\
-\ \frac{\eps^3}{2\mu_1(\eps)}\
\Delta_{(X_1,X_2,X_3)}\ -\
\frac{\eps^4}{2\mu_2(\eps)}\
\Delta_{(Y_1,Y_2,Y_3)}\ +\
h(\eps,\,X_1,\,X_2,\,X_3,\,Y_1,\,Y_2,\,Y_3),
$$
rewritten in terms of the variables
$(Y,\,\theta,\,\phi,\,R,\,\gamma,\,X,\,Z)$.

The quasienergy will be
\be\label{horse}
{\cal E}(\eps)\ =\
{\cal E}_0\ +\
\eps^{6/4}\ {\cal E}_{6/4}\ +\
\eps^{8/4}\ {\cal E}_{8/4}\ +\
\eps^{10/4}\ {\cal E}_{10/4},
\ee
but the quasimode will be somewhat complicated.

To specify the quasimode, we first let
$\chi(Y,\,\theta,\,\phi,\,R,\,\gamma,\,X,\,Z)$
denote a normalized real ground state eigenfunction
of $h(\eps,\,Y,\,\theta,\,\phi,\,R,\,\gamma,\,X)$
that depends continuously on its variables.
Next, we let
$$
\zeta(\eps,\,Y,\,\theta,\,\phi,\,R,\,\gamma,\,X)\ =\
\eps^{-5/4}\ 
\sum_{l=0}^5\ \eps^{l/4}\
\psi_{l/4}\left(\,\frac{X-X_0}{\eps^{3/4}},\,
\frac R{\eps^{1/2}},\,\frac{Y-Y_0}{\eps^{3/4}},
\,\theta,\,\phi,\,\gamma\right),
$$
where the $\psi_{l/4}$ are the wave functions from
Section \ref{Sect3} with
$g_{2/4}(y,\,\pm\lambda)=g_{3/4}(y,\,\pm\lambda)=
f_{4/4}(r,\,y,\,k)=f_{5/4}(r,\,y,\,k)=0.$
Note that when $\lambda=0$ there is one linearly
independent choice for $\zeta$. When $\lambda>0$, we
have two linearly independent choices corresponding
to $k=\pm\lambda$.

The quasimode is
\bea\nonumber
&&\hskip -17mm
\Psi(\eps,\,Y,\,\theta,\,\phi,\,R,\,\gamma,\,X,\,Z)
\\[3mm]\nonumber
&&\hskip -17mm =\quad
{\cal F}_\eps(X,\,R,\,Y)\
\zeta(\eps,\,Y,\,\theta,\,\phi,\,R,\,\gamma,\,X)\
\chi(Y,\,\theta,\,\phi,\,R,\,\gamma,\,X,\,Z)
\\[3mm]\label{quasimodo1}
&&\hskip -7mm +\quad
\frac{\eps^3}{2\,\mu_1}\ {\cal F}_\eps(X,\,R,\,Y)\
\Big[\,h(\eps,\,Y,\,\theta,\,\phi,\,R,\,\gamma,\,X)
\,-\,V(\eps,\,X,\,R,\,Y)\,\Big]_r^{-1}
\\[3mm]\nonumber
&&\hskip 1cm \times\quad\left(\,
\frac{\partial\zeta}{\partial X}
(\eps,\,Y,\,\theta,\,\phi,\,R,\,\gamma,\,X)\
\frac{\partial\chi}{\partial X}
(Y,\,\theta,\,\phi,\,R,\,\gamma,\,X,\,Z)
\right.
\\[3mm]\nonumber
&&\hskip 45mm +\quad
\left.\frac{\partial\zeta}{\partial R}
(\eps,\,Y,\,\theta,\,\phi,\,R,\,\gamma,\,X)\
\frac{\partial\chi}{\partial R}
(Y,\,\theta,\,\phi,\,R,\,\gamma,\,X,\,Z)
\right).
\eea
%We note that this function has norm
%$1\,+\,O(\eps^{1/2})$ since $\psi_0$ has norm
%$1$, $\psi_{j/4}$ is orthogonal to $\psi_0$ for
%each $j$, and $\psi_{1/4}$ is non-trivial.

\vskip 5mm
\begin{theorem}\label{theorem2}
There exists a constant $C$, such that the function
$\Psi(\eps)$ given by (\ref{quasimodo1})
and quasienergy ${\cal E}(\eps)$ given by
(\ref{horse}) satisfy
$$
\|\,\Psi(\eps)\,\|\ =\
1\ +\ O\left(\eps^{1/2}\right)
$$
and
\be\label{moose}\hspace{4cm}
\left\|\,\Big(\,H(\eps)\,-\,{\cal E}(\eps)\,\Big)\
\Psi(\eps,\,\cdot)\,\right\|
\ \le\ C\ \eps^3
\ee
\end{theorem}

\vskip 5mm \noindent
{\bf Proof}\quad
The function~ $\Psi(\eps,\,\cdot\,)$~ equals
the normalized vector~ $\psi_0\ \chi$~
plus terms that are orthogonal to $\psi_0\,\chi$.
Since the largest of these orthogonal terms is~
$\eps^{1/4}\,\psi_{1/4}\,\chi$,~ we see that~
$\Psi(\eps)$~ has norm~ $1\,+\,O(\eps^{1/2})$.

To prove the second estimate of the theorem, we
begin by noting that the electronic
eigenfunction $\chi$ has the form
$$
\chi(Y,\,\theta,\,\phi,\,R,\,\gamma,\,X,\,Z)\ =\
U(\theta,\,\phi,\,\gamma)\
\chi_0(Y,\,R,\,X,\,Z),
$$
where $U(\theta,\,\phi,\,\gamma)$ is unitary.

We next compute
\be\label{mainterm}
\Big(\,H(\eps)\,-\,{\cal E}(\eps)\,\Big)\
{\cal F}_\eps(X,\,R,\,Y)\ \zeta(\eps,\,\cdot)\ \chi(\cdot),
\ee
where $H(\eps)$ is decomposed as
\bea\nonumber
H(\eps)&=&
-\ \frac{\eps^3}{2\mu_1(\eps)}\
\Delta_{(X_1,X_2,X_3)}\ -\
\frac{\eps^4}{2\mu_2(\eps)}\
\Delta_{(Y_1,Y_2,Y_3)}
\\[3mm]\nonumber
&&+\quad\Big[\,
h(\eps,\,X_1,\,X_2,\,X_3,\,Y_1,\,Y_2,\,Y_3)\ -\
V_1(X)\ -\ \eps\ V_2(X,\,R,\,Y)\,\Big]
\\[3mm]\nonumber
&&+\quad
V_1(X)\ +\ \eps\ V_2(X,\,R,\,Y),
\eea
with the two final terms expanded in their Taylor
series of appropriate orders.
We write the resulting expression in the variables
$(Y,\,\theta,\,\phi,\,R,\,\gamma,\,X,\,Z)$.
When the all the derivatives in $H(\eps)$
act on $\zeta$, all terms that are larger than
order $\eps^3$ cancel because of Taylor series estimates
and the choices of the $\psi_{l/4}$.
When all the derivatives act on $\chi$, all terms are
$O(\eps^3)$ or smaller because $\chi$ is smooth and
the cutoffs are zero the singularity at $Y=0$.
When any derivatives act on ${\cal F}_\eps$, we obtain
terms of order $O(\eps^q)$, for any $q$,
due to the rapid fall off of the functions in $\zeta$.
The term that arises from~
$[h(\eps)-V_1-\eps\,V_2]$~ yields zero because it acts
only on the $\chi$.

The remaining terms in (\ref{mainterm}) contain terms
in which a partial derivative acts on $\zeta$ and the
same partial derivative acts on $\chi$.
All of these terms are $O(\eps^3)$ or
smaller, except for
\bea\label{cow}
&&\frac{\partial\zeta}{\partial X}
(\eps,\,Y,\,\theta,\,\phi,\,R,\,\gamma,\,X)\
\frac{\partial\chi}{\partial X}
(Y,\,\theta,\,\phi,\,R,\,\gamma,\,X,\,Z)
\\[3mm]\nonumber
&+&
\frac{\partial\zeta}{\partial R}
(\eps,\,Y,\,\theta,\,\phi,\,R,\,\gamma,\,X)\
\frac{\partial\chi}{\partial R}
(Y,\,\theta,\,\phi,\,R,\,\gamma,\,X,\,Z).
\eea
Thus, (\ref{mainterm}) yields
(\ref{cow}) plus $O(\eps^3)$. However, when the
$[h(\eps)-V_1-\eps\,V_2]$ acts on the second term
in (\ref{quasimodo1}), the terms that arise from (\ref{cow})
cancel, leaving us with $O(\eps^3)$ errors plus
the kinetic energy and potential terms acting on
the second term in (\ref{horse}). Because of the 
cutoff, the potential terms yield bounded operators
times $O(\eps^3)$ terms. When the kinetic energy
acts on these terms, we obtain terms of order
$\eps^{9/2}$ or smaller, since everything is smooth,
and the largest terms come from $\eps^6$ and
two $X$--derivatives acting on $\zeta$.

Note that when computing the norm in 
(\ref{moose}), it is essential that
$\chi$ be orthogonal to~
$\ds\frac{\partial\chi}{\partial X}$
and~
$\ds\frac{\partial\chi}{\partial R}$,~
or cross terms would yield terms of order
greater than $\eps^3$.
This orthogonality is guaranteed by our
hypothesis that the electron Hamiltonian
$h(\eps,\,\cdot\,)$ be real symmetric
and that we choose $\chi$ to be real.
\hfill\ep

\end{document}